  \newcommand{\sgn}{\operatorname{sgn}}

\documentclass[11pt, a4paper]{article}

\usepackage[table]{xcolor}

\usepackage{array}

\usepackage{float}

\usepackage{amsmath,amssymb}

\usepackage{changepage}

\usepackage[utf8x]{inputenc}

\usepackage{textcomp,marvosym}

\usepackage{cite}

\usepackage[english]{babel}
\usepackage[utf8x]{inputenc}
\usepackage[T1]{fontenc}

\usepackage[utf8x]{inputenc}

\usepackage{textcomp,marvosym}

\usepackage{cite}

\usepackage[a4paper,top=3cm,bottom=2cm,left=3cm,right=3cm,marginparwidth=1.75cm]{geometry}
\usepackage{amsmath}
\usepackage{graphicx}
\usepackage[colorinlistoftodos]{todonotes}
\usepackage[colorlinks=true, allcolors=blue]{hyperref}

\title{Learning Mid-Level Auditory Codes from Natural Sound Statistics}
\author{Wiktor M\l ynarski*, Josh H. McDermott\\\small{mlynar@mit.edu}}
\date{}

\begin{document}
\maketitle

\begin{abstract}
Interaction with the world requires an organism to transform sensory signals into representations in which behaviorally meaningful properties of the environment are made explicit. These representations are derived through cascades of neuronal processing stages in which neurons at each stage recode the output of preceding stages. Explanations of sensory coding may thus involve understanding how low-level patterns are combined into more complex structures. To gain insight into such mid-level representations for sound, we designed a hierarchical generative model of natural sounds that learns combinations of spectrotemporal features from natural stimulus statistics. In the first layer the model forms a sparse convolutional code of spectrograms using a dictionary of learned spectrotemporal kernels. To generalize from specific kernel activation patterns, the second layer encodes patterns of time-varying magnitude of multiple first layer coefficients. When trained on corpora of speech and environmental sounds, some second-layer units learned to group spectrotemporal features that occur together in natural sounds. Others instantiate opponency between dissimilar sets of spectrotemporal features. Such groupings might be instantiated by neurons in the auditory cortex, providing a hypothesis for mid-level neuronal computation. \end{abstract}

\section*{Introduction}

The ability to interact with the environment requires an organism to infer characteristics of the world from sensory signals. One challenge is that the environmental properties an organism must recognize are usually not explicit in the sensory input. A primary function of sensory systems is to transform raw sensory signals into representations in which behaviorally important features are more easily recovered. In doing so the brain must generalize across irrelevant stimulus variation while maintaining selectivity to the variables that matter for behavior. The nature of sensory codes and the mechanisms by which they achieve appropriate selectivity and invariance are thus a primary target of sensory system research.
 
The auditory system is believed to instantiate such representations through a sequence of processing stages extending from the cochlea into the auditory cortex. Existing functional evidence suggests that neurons in progressively higher stages of the auditory pathway respond to increasingly complex and abstract properties of sound \cite{Chechik, AtencioExpands, CarruthersGeffen, BizleyKingNelken, ElieTheunissen, RussCohen, MesgeraniFritz, ObleserLeaver, BendorWang, Overath, NormanHaignere}. Yet our understanding of the underlying transformations remains limited, particularly when compared to the visual system. 

Feature selectivity throughout the auditory system has traditionally been described using linear receptive fields \cite{Aertsen, TheunissenSenDoupe, Shamma}. The most common instantiation is the spectrotemporal receptive field (STRF), which typically characterizes neural activity with a one-dimensional linear projection of the sound spectrogram transformed with a nonlinearity \cite{SharpeeHier}. As a neural data analysis technique, STRFs are widespread in auditory neuroscience and have generated insight in domains ranging from plasticity to speech coding (e.g. \cite{ShammaTaskSpecific, WoolleyTheunissen}). 

Despite their utility, it is clear that STRFs are at best an incomplete description of auditory codes, especially in the cortex \cite{Machens, Sahani, Williamson}. Experimental evidence suggests that auditory neural responses are strongly nonlinear. As a consequence, auditory receptive fields estimated with natural sounds differ substantially from estimates obtained with artificial stimuli \cite{TheunissenSenDoupe}. STRF descriptions also fail to capture the dimensionality expansion of higher representational stages. In contrast to the brainstem, neurons in the auditory cortex seem to be sensitive to multiple stimulus features at the same time \cite{AtencioExpands, KozlovGentnerComposite, Harper}. The presence of strongly non-linear behavior and multiplexing necessitates signal models more sophisticated than one-dimensional, linear features of the spectrogram such as STRFs.

An additional challenge to characterizing mid-level features of sound is that humans lack strong intuitions about abstract auditory structure. In specific signal domains such as speech, progress has been made by cataloging phonemes and other frequently occurring structures, but it is not obvious how to generalize this approach to broader corpora of natural sounds. 

An alternative approach to understanding sensory representations that is less reliant on domain-specific intuition is that of efficient coding \cite{Barlow, Attneave}. The efficient coding hypothesis holds that neural codes should exploit the statistical structure of natural signals, allowing such signals to be represented with a minimum of resources. Numerous studies have demonstrated that tuning properties of neurons in early stages of the visual and auditory systems are predicted by statistical models of natural images or sounds \cite{Srinivasan, OlshausenField, BellSejnowski, vanHateren,Lewicki, klein2003sparse, SmithLewicki, Carlson,TerashimaOkada, Mlynarski, Mlynarski2, Elhilali,DeepBeliefV2,HyvarinenHosoya}. Although the early successes of this approach engendered optimism, applications have largely been limited to learning a single stage of representation, and extensions to multiple levels of sensory processing have proven difficult. The underlying challenge is that there are many possible forms of high-order statistical dependencies in signals, and the particular dependencies that occur in natural stimuli are typically not obvious. The formulation of models capable of capturing these dependencies requires careful analysis and design \cite{HyvarinenISA,KarklinVariance,Karklin,Olshausen}, and perhaps good fortune, and is additionally constrained by what is tractable to implement. In the auditory system in particular, it remains to be seen whether modeling statistical signal regularities can reveal the complex acoustic structures and invariances that are believed to be represented in higher stages of the auditory system.

The primary goal of the present work was to discover such high-order structure in natural sounds and generate hypotheses about not-yet-observed intermediate-level neural representations. To this end we developed a probabilistic generative model of natural sounds designed to learn a novel stimulus representation - a population code of naturally occurring combinations of basic spectrotemporal patterns, analogous to STRFs. 

The representations learned by the model from corpora of natural sounds suggest grouping principles in the auditory system. Model units learned to pool sets of similar spectrotemporal features, presumably because they occur together in natural sounds. In addition, some model units encode opponency between different sets of features. Although not yet described in the auditory system, such tuning patterns may be analogous to phenomena such as end-stopping or cross-orientation suppression in the visual system.  The representations learned by our model also resemble some recently reported properties of auditory cortical neurons, providing further evidence that natural-scene statistics can predict neural representations in higher sensory areas.

\section*{Methods and Models}
\subsection*{Overview of the hierarchical model}

\begin{figure}[H]
  \centering
  \includegraphics[scale=1]{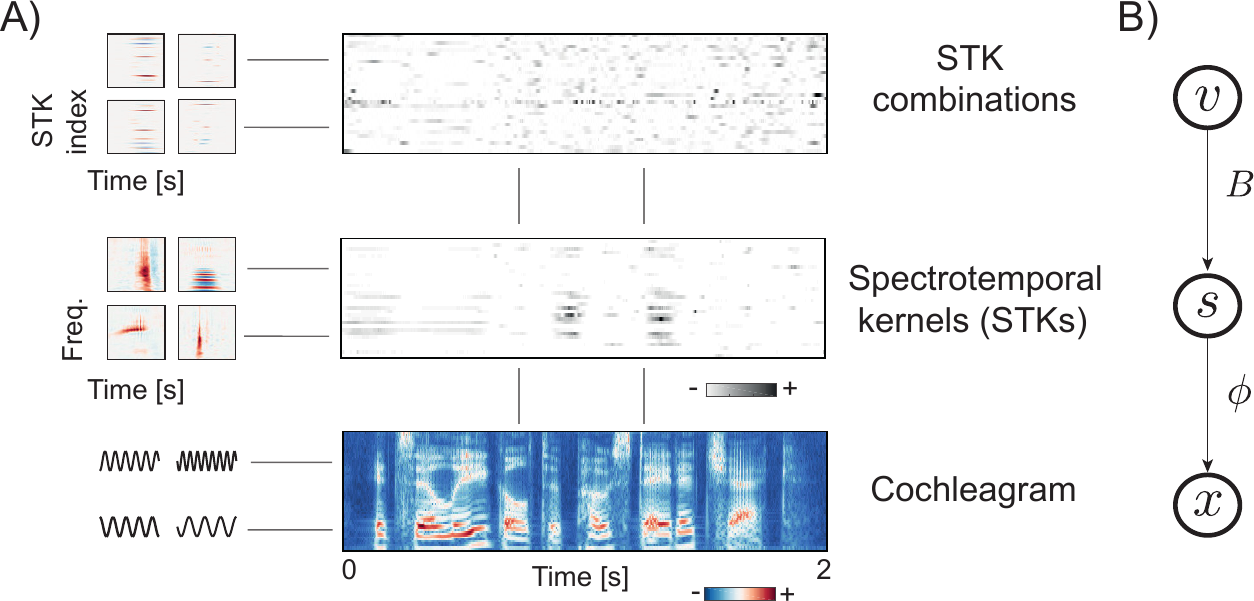}
  \caption{{\bf Overview of the hierarchical model.} A) A spectrogram (bottom-row) is encoded by a set of spectrotemporal kernels (middle row). The features learned by the second layer encode temporal patterns of multiple STK activations. B) A graphical model depicting statistical dependencies among variables.}
\label{fig1Scheme} 
\end{figure}

To learn mid-level auditory representations, we constructed a hierarchical, statistical model of natural sounds. The model structure is depicted in Fig \ref{fig1Scheme}. The model consisted of a stimulus layer and two latent layers that were adapted to efficiently represent a corpus of audio signals. Because our goal was to learn mid-level auditory codes, we did not model the raw sound waveform. Instead, we assumed an initial stage of frequency analysis, modeled after that of the mammalian cochlea. This frequency analysis results in a spectrogram-like input representation of sound, which we term a 'cochleagram,' that provides a coarse model of the auditory nerve input to the brain (Fig \ref{fig1Scheme} A, bottom row). This input representation is an $F \times T$ matrix, where $F$ is the number of frequency channels and $T$ the number of time-points. Our aim was to capture statistical dependencies in natural sounds represented in this way.

The first layer of the model (Fig \ref{fig1Scheme}, middle row) was intended to learn basic acoustic features, analogous to STRFs of early auditory neurons. Through the remainder of the paper, we refer to the features learned by the first layer as spectrotemporal kernels (STKs), in order to differentiate them from neurally derived STRFs. The second layer, depicted in the top row of Fig \ref{fig1Scheme} A, was intended to learn patterns of STK co-activations that frequently occur in natural sounds.

The model specifies a probability distribution over the space of natural sounds, and its parameters can be understood as random variables whose dependency structure is depicted in Fig \ref{fig1Scheme} B. The spectrogram $x$ is represented with a set of spectrotemporal features $\phi$ convolved with the latent activation time-courses $s$. The second latent layer encodes the magnitudes of $s$ with the basis functions $B$ convolved with their activation time-courses $v$. In the following sections, we present the details of each layer.

\subsection*{First layer of model - convolutional, non-negative sparse coding of spectrograms}

The first layer of the model was designed to learn basic spectrotemporal features. One previous attempt to learn sparse spectrotemporal representations of natural sounds \cite{Carlson} produced structures reminiscent of receptive fields of neurons in the auditory midbrain, thalamus, and cortex. Here, we extended this approach by learning a spectrogram representation which is sparse and convolutional.

\begin{figure}[H]
  \centering
  \includegraphics[scale=1]{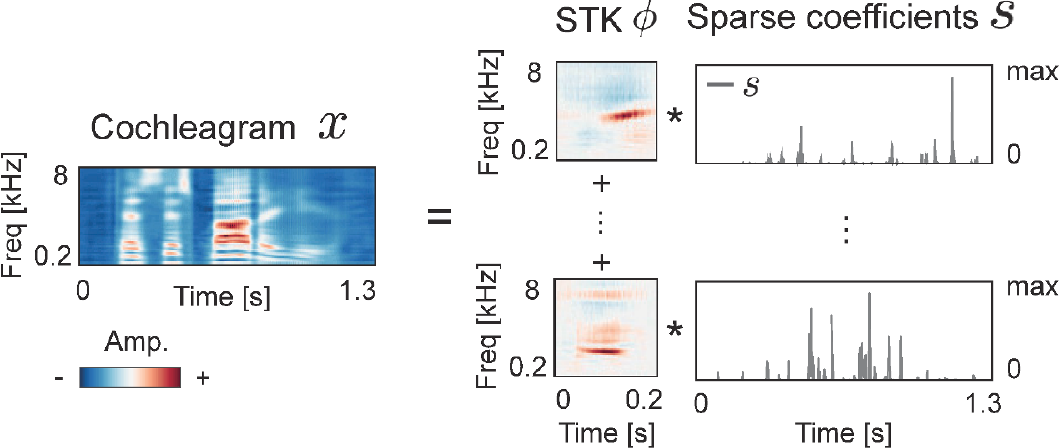}
  \caption{{\bf Explanation of first layer of the model.} A cochleagram $x$ of arbitrary length is represented as a sum of spectrotemporal kernels $\phi$ convolved with time-courses of corresponding coefficients $s$. Kernels span the full extent of the frequency dimension, and convolution occurred only in time. Coefficients $s$ are non-negative and have sparse distributions (i.e. remain close to zero most of the time).}
\label{figFirstLayer} 
\end{figure}

A schematic of the first layer is depicted in Fig \ref{figFirstLayer}. We modeled the cochleagram  ($x_{t,f}$) as a linear combination of spectrotemporal kernels $\phi$ convolved in time with their activation time-courses $s$ and distorted by additive Gaussian noise $\xi_{t,f}$ with variance $\sigma^2$.:
\begin{align}
\label{eq1:scc}
    \hat{x}_{t,f} &= \left[\sum_{i=1}^N \phi_{i,f} \ast s_i\right]_t \\
	x_{t,f} &= \hat{x}_{t,f} + \xi_{t,f}
\end{align}

Kernel activations $s_{i,t}$ are assumed to be independent, i.e. their joint distribution is equal to the product of marginals:
\begin{align}
\label{eq2:sCoeffs}
	p(\mathbf{s}) = \prod_{i=1}^N \prod_{t=1}^T p(s_{i,t} | \lambda_i)
\end{align}

We assumed that each spectrotemporal kernel remains inactive for most of the time, i.e. that the distribution of its activations is sparse. Moreover, we imposed a non-negativity constraint on the coefficients $s$. This facilitates interpretations in terms of neural activity and improves the interpretability of the learned representation. These constraints were embodied in an exponential prior on the coefficients $s$:
\begin{equation}
\label{eq3:exponential}
p(s_{i,t}|\lambda_i) = \frac{1}{\lambda_i} \exp \left[-\frac{s_{i,t}}{\lambda_i}\right]
\end{equation}
where $\lambda_i$ is the scale parameter. The scale of the exponential distribution determines its "spread" along the real line. Small values of $\lambda_i$ yield distributions tightly peaked at $0$. Large $\lambda_i$ generate broader distributions allowing $s_{i,t}$ to attain large values with higher probability. When training the first layer we assumed all $\lambda_i$ to be constant and equal to $1$. As we will describe in the following sections, the second layer of the model relaxes that assumption and learns a representation of time-varying scale parameters $\lambda_i$.

The first layer of the model specifies the following negative log-posterior probability of the data:
\begin{equation}
\label{eq3:E1}
	E_1 \propto \frac{1}{\sigma^2} \sum_{f=1}^F\sum_{t=1}^T (\hat{x}_{t,f} - x_{t,f})^2 + \sum_{i=1}^{N}\frac{1}{\lambda_i}\sum_{t=1}^T  s_{i,t}
\end{equation}
This negative log-probability can be viewed as a cost function to be minimized when inferring the value of coefficients $s$: while maintaining low-reconstruction error (first term on the right-hand side), the sparsity of representation should be maximized (the second term). 

\subsubsection*{Dependencies between coefficients - a signature of mid-level structure}
\label{SectionDependencies}

Although the sparse coding strategy outlined above learns features that are approximately independent across the training set, residual dependencies nonetheless remain. In part this is because not all dependencies can be modeled with a single layer of convolutional sparse coding. However, due to the non-stationary nature of natural audio, coefficient dependencies can also be present locally despite not being evident across a large corpus as a whole. Empirically, the learned features exhibit dependencies specific to particular sounds \cite{Karklin}, and thus deviate locally from their (approximately independent) marginal distribution. For example, a spoken vowel with a fluctuating pitch contour would require many harmonic STKs to become activated, and their activations would become strongly correlated on a local time scale. Such local correlations reflect higher-order structure of particular natural sounds. Statistically speaking, this is an example of marginally independent random variables exhibiting conditional dependence (in this case conditioned on a particular point in time or a type of sound). 

\begin{figure}
  \centering
  \includegraphics[scale=1]{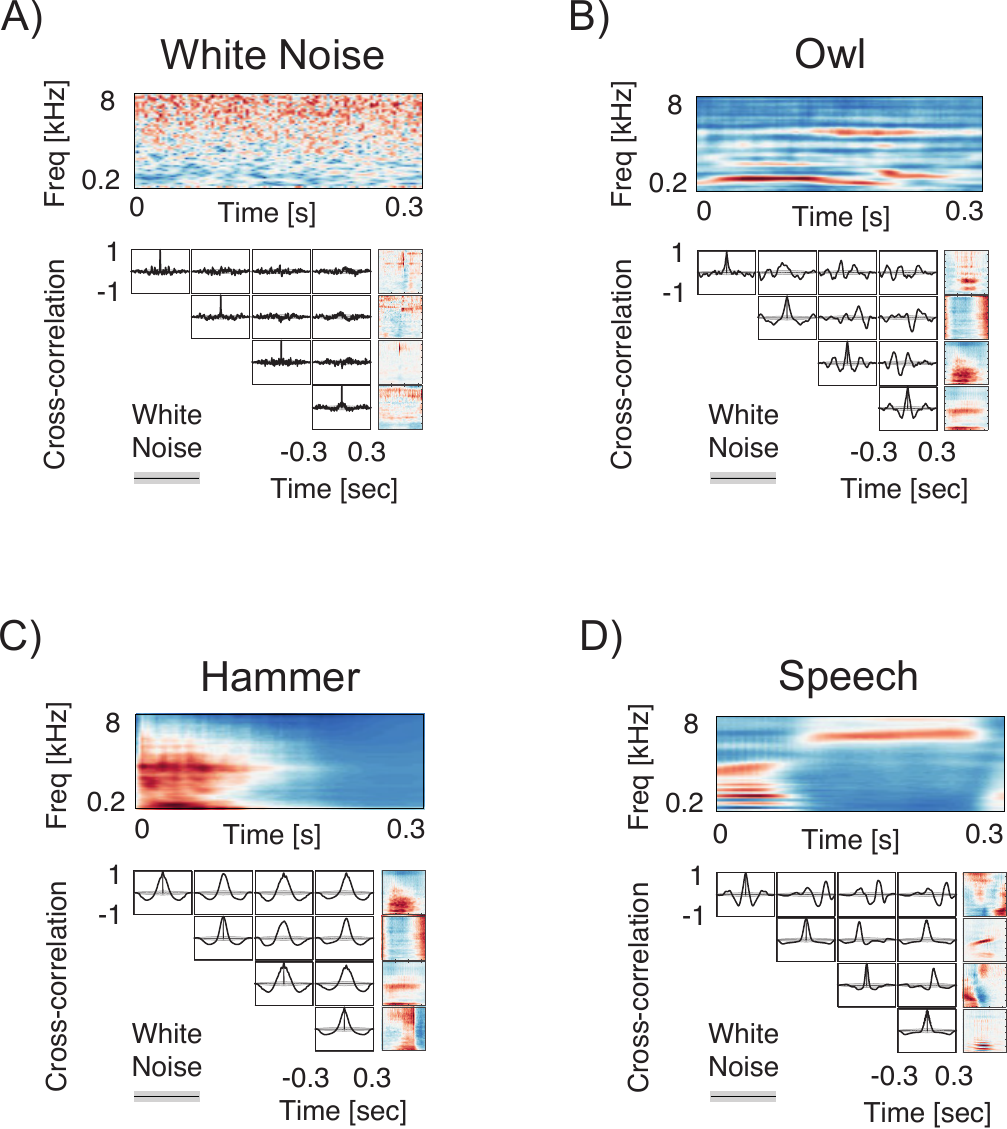}
  \caption{{\bf Dependencies among spectrotemporal feature activations.} A) When encoding white noise (cochleagram depicted on top), sparse coefficients are uncorrelated on short time-scales. This is visible in the shape of the coefficient cross-correlation functions (black lines) corresponding to the four most strongly activated STKs (bottom, right column). Due to the lack of temporal structure, the cross-correlation between STKs is flat, and the autocorrelation of individual STKs is a Dirac delta function at $0$. Here and in other panels, the mean was substracted from the (non-negative) coefficient trajectories prior to computing the cross-correlation. B) When encoding structured stimuli such as an owl vocalization, STK activations reveal strong local correlations - the cross-correlations deviate from those for white noise (thick gray lines). C) Same as B, for a hammer hit. D) Same as B, for a speech excerpt. }
\label{fig3Dependencies} 
\end{figure}

Fig \ref{fig3Dependencies} depicts such dependencies via cross-correlation functions of selected STK activations for a white noise sample and for three different natural sounds. Coefficient correlations (black curves in each subplot) vary from sound to sound, but in all cases deviate from those obtained with noise (gray bars within subplots), revealing dependence. These dependencies are indicative of "mid-level" auditory features, perhaps analogous to the correlations between oriented Gabor filters induced by an elongated edge. In this work, we exploited the fact that intermediate level representations can be learned by modeling dependencies among first-layer features \cite{Karklin, KarklinVariance, Olshausen}.

\subsubsection*{Variability of STK activations}
\label{SectionAmplitudes}

\begin{figure}[H]
  \centering
  \includegraphics[scale=1]{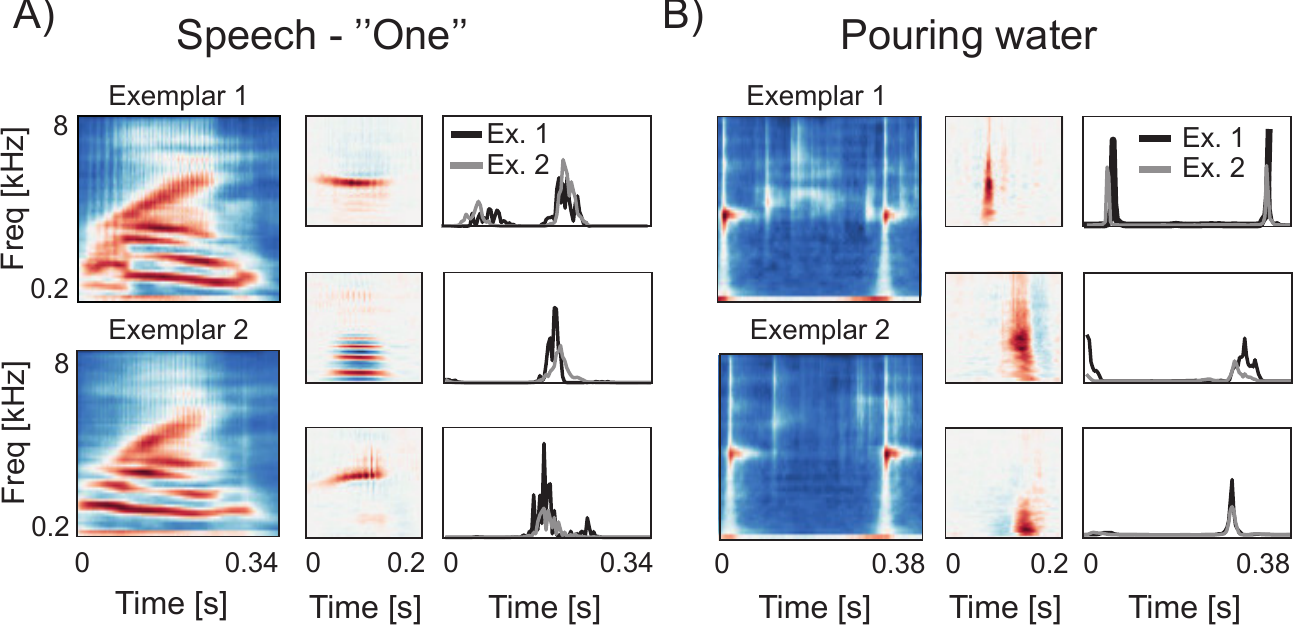}
  \caption{{\bf Variability of spectrotemporal feature activations.} A) Coefficient trajectories of spectrotemporal features $\phi$ are depicted for two utterances of word "one" spoken by the same speaker. Although both coefficient trajectories (gray and black lines) exhibit the same global structure, they are not identical. B) Coefficient variability visualized in the same way for two different examples of water pouring into a cup.}
\label{fig4Variance} 
\end{figure}

A second phenomenon evident in STK activations is that particular patterns of co-activation occur with some variability. This is visible in Fig \ref{fig4Variance}, which depicts activations of selected STKs when encoding multiple exemplars of the same sound - the word "one" spoken twice by the same speaker (Fig \ref{fig4Variance}A) and two exemplars of water being poured into a cup (Fig \ref{fig4Variance}B).

It is apparent that the STK coefficient trajectories for the two exemplars in each case (black and gray lines) reflect the same global pattern even though they differ somewhat from exemplar to exemplar. The similarity suggests that the trajectories could be modeled as different samples from a single time-varying distribution parameterized by a non-stationary coefficient magnitude. When the magnitude increases, the probability of a strong STK activation increases. Retaining the (inferred) time-varying magnitude instead of precise values of STK coefficients would yield a representation more invariant to low-level signal variation, potentially enabling the representation of abstract regularities in the data. Such a representation bears an abstract similarity to the magnitude operation used to compute a spectrogram, in which each frequency channel retains the time-varying energy in different parts of the spectrum. Here we are instead estimating a scale parameter of the underlying distribution, but the process similarly discards aspects of the fine detail of the signal. 

\subsection*{Second layer of model - encoding of STK combinations}

The second layer of the model was intended to exploit the two statistical phenomena detailed in the previous section: conditional dependencies between STKs and their variation across exemplars. Similarly to the first layer, the second layer representation is formed by a population of sparsely activated basis functions. These basis functions capture local dependencies among STK magnitudes by encoding the joint distribution of STK activations rather than exact values of STK coefficients. The resulting representation is thus more specific than the first-layer code - instead of encoding single features independently, it signals the presence of particular STK combinations. It is also more invariant, generalizing over specific coefficient values.

Because the proposed representation is a population code of a distribution parameter, it bears conceptual similarity to previously proposed hierarchical models of natural stimuli that encoded patterns of variance \cite{KarklinVariance, BumbacherMing}, covariance \cite{Karklin} or complex amplitude \cite{Olshausen,HyvarinenISA}. The novelty of our model structure lies in being convolutional (i.e., it can encode stimuli of arbitrary length using the same representation) and in parameterizing distributions of non-negative STK coefficients, increasing the interpretability of the learned spectrogram features. The novelty of the model's application is to learn hierarchical representations of sound (previous such efforts have largely been restricted to modeling images; though see \cite{YNg}).

\begin{figure}[H]
  \centering
  \includegraphics[scale=1]{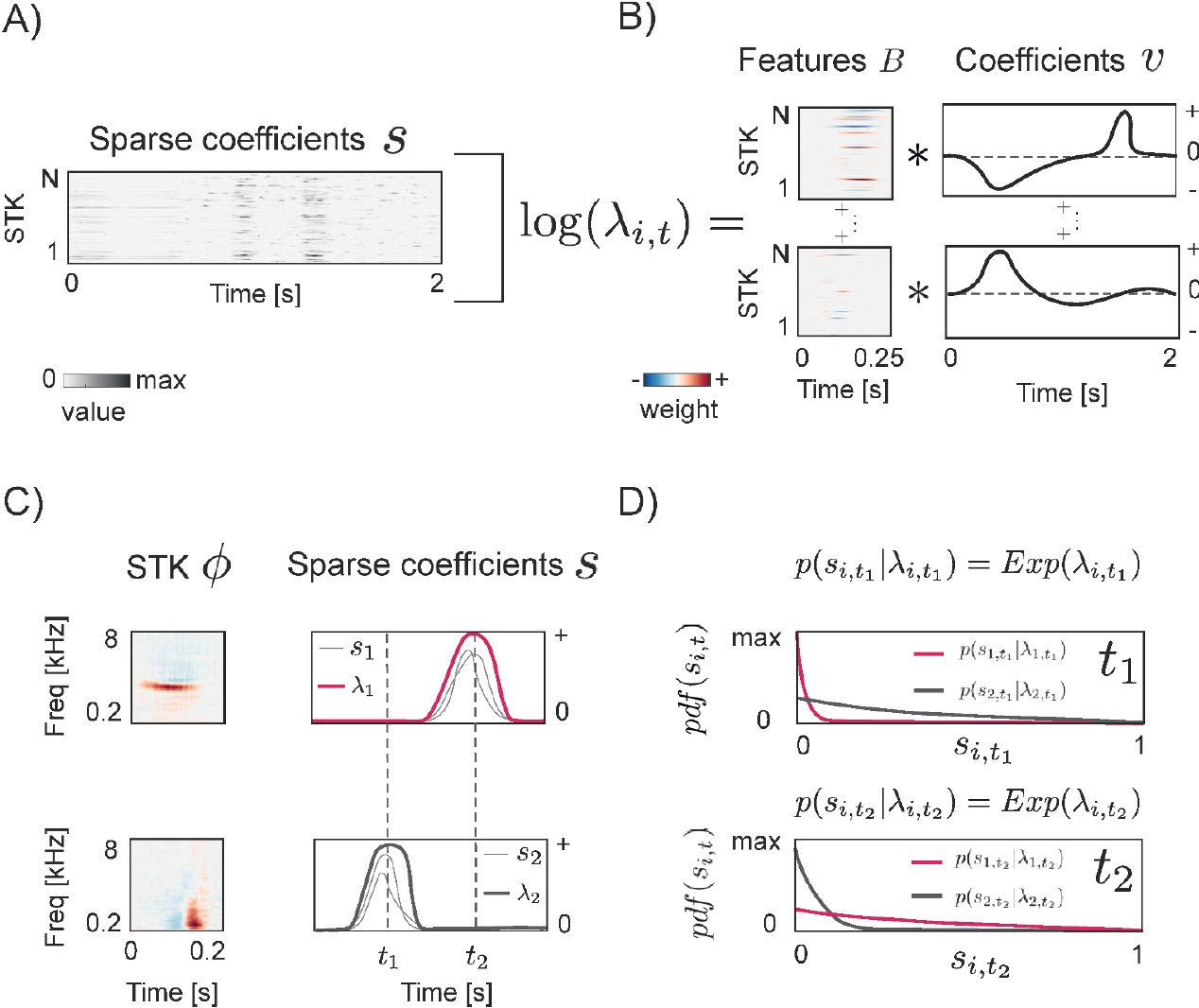}
  \caption{{\bf Explanation of second layer of the model.} A) An array of STK activations $s$ (a "STK-gram") serves as an input to the second layer. Rows correspond to first layer features $\phi_i$ and columns to time points. B) The second layer uses a population of features $B$ to encode the logarithm of STK activation magnitudes. C) Coefficient trajectories $s$ (thin grey lines) and their magnitudes $\lambda$ (thick red and black lines) for two example STKs. D) Distributions of $s$ at time points $t_1$ and $t_2$ are depicted in the right column.}
\label{fig5Layer2Explanation} 
\end{figure}

The second layer of the model is depicted schematically in Fig \ref{fig5Layer2Explanation}A. We assume that STK activations $s$ are samples from a non-stationary exponential distribution with time-varying scale parameter $\lambda_{i,t}$, relaxing the assumption of stationary $\lambda_i$ made in learning the first layer:

\begin{equation}
\label{eq4:lambda}
	p(s_{i,t}) = Exp(\lambda_{i,t}) = \frac{1}{\lambda_{i,t}}\exp\left[-\frac{s_{i,t}}{\lambda_{i,t}} \right]
\end{equation}

When $\lambda_{i,t}$ is high, the distribution of $s_{i,t}$ becomes broader (Fig \ref{fig5Layer2Explanation}C, black line in the top row, red line in the bottom). This allows the coefficient $s_i$ to attain large values. For small values of the scale parameter, the probability density is concentrated close to $0$ (Fig \ref{fig5Layer2Explanation}D), red line in the top row, black line in the bottom), and coefficients $s_{i,t}$ become small. To model the magnitudes $\lambda$  (which are non-negative, akin to variances), we took their logarithm, mapping their values onto the entire real line so that they could be represented by a sum of real-valued basis functions. 

Patterns of STK magnitudes are represented in the second layer by a population of features $B$ convolved with coefficients $v$:
\begin{equation}
\label{eq5:layer2}
	\lambda_{i,t} = \exp \left[\sum_{j=1}^M B_{j,i} \ast v_j + \rho_i \right]_t
\end{equation}

where $\rho$ is a bias vector. Each second-layer basis function $B$ represents a particular temporal pattern of co-activation of first-layer STKs. Their corresponding coefficients $v$ are assumed to be sparse and independent:

\begin{align}
\label{eq6:vCoeffs}
	p(\bf{v}) &= \prod_{j=1}^M \prod_{t=1}^T p(v_{j,t})\\
	p(v_{j,t}) &\propto \exp\Big(-\alpha |v_{j,t}|\Big)
\end{align}

where $\alpha$ controls the degree of sparsity.

As with the first layer, learning and inference are performed by gradient descent on the negative log posterior. Because the second layer units encode combinations of sparse first-layer coefficients, we placed a sparse prior on the $L_1$ norm of the basis functions $B$. The overall cost function to be minimized during learning in the second layer is then:

\begin{equation}
\label{eq7:E2}
	E_2 \propto \sum_{i=1}^N\sum_{t=1}^T \frac{s_{i,t}}{\lambda_{i,t}} + \log(\lambda_{i,t}) + \alpha \sum_{j=1}^M\sum_{t=1}^T |v_{j,t}| + \beta \sum_{j=1}^M\sum_{i=1}^N\sum_{t_b=1}^{T_b} |B_{j,i,t_b}|
\end{equation}

where $\beta$ controls the strength of the sparse prior on $B$, and $T_b$ is the temporal extent of each second-layer basis function. As in the first layer cost function (Eq \ref{eq3:E1}), the first term on the right hand side of Eq \ref{eq7:E2} enforces a match of the representation to data (the magnitudes $\lambda$ are pushed away from zero towards the observed coefficients $s$), while the second and third terms promote sparsity of second-layer coefficients and basis functions, respectively. Sparsity of second layer units is a reasonable assumption given the sparsity of first-layer coefficients, and we found that using a sparse prior on second-layer units (i.e. setting $\beta$ to be larger than $0$) was necessary to achieve convergence. Others have found that the addition of sparse priors can enable faster learning with less data without substantially altering the obtained solution \cite{RajuHyvarinen}. 

\subsection*{Learning procedure}

The two layers of the model were trained separately, i.e. the training of the second layer occurred after the first layer training was completed. In the first layer, training was performed with an EM-like procedure that iteratively alternated between inferring STK coefficients and updating STK features \cite{OlshausenField,Olshausen}. Spectrotemporal features $\phi$ were initialized with Gaussian white noise. For each excerpt in the training set, coefficients $s_{i,t}$ were inferred via gradient descent on the energy function (\ref{eq3:E1}). Because inference of all coefficients $s$ is computationally expensive, we adopted an approximate inference scheme \cite{ConvSparse}. Instead of inferring values of all coefficients for each excerpt, we selected only a subset of them to be minimized. This was done by computing the cross-correlation between a sound excerpt and features $\phi_i$ and selecting a fixed number of the largest coefficients $s_{i,t}$. The inference step adjusted only this subset of coefficients while setting the rest to $0$. Given the inferred coefficients, a gradient step on the spectrotemporal features $\phi$ was performed.

Each learning iteration therefore consisted of the following steps:
\begin{enumerate}
	\item Draw a random sound excerpt from the training data set. Excerpts were $403$ ms in length ($129$ time samples of the spectrogram, sampled at $320$ Hz).
    
    \item Compute the cross-correlation of all basis functions $\phi_i$ with the sound excerpt. Select the $1024$ pairs of coefficient indices and time-points $(i, t)$ that yield the highest correlation values.
    
    \item Infer the values of the selected coefficients by minimizing Eq. \ref{eq3:E1} with respect to $s_{i,t}$ via gradient descent. Set the rest of coefficients to $0$.
    
   	\item Compute the gradient step on the basis functions as the derivative of Eq. \ref{eq3:E1} with respect to basis functions $\phi$ using inferred coefficient values $\hat{s}$. Update basis functions according to the gradient step.
    
    \item Normalize all basis functions to unit norm.
\end{enumerate}

This procedure was terminated after $200000$ iterations.

The second layer was then learned via the same procedure used for the first layer. In each iteration a $528$ ms long ($169$ samples at $320$ Hz) randomly drawn sound excerpt was encoded by the first layer, and the resulting matrix of coefficients $s$ served as an input to the second layer. A subset of coefficients $v$ was selected for approximate inference by computing the cross correlation between features $B_i$ and the logarithm of the first-layer coefficients $s$ (analogous to step $2$ in the procedure described above for the first layer). The energy function $E_2$ was first minimized with respect to coefficients $v$ followed by a gradient update to the basis functions $B$ (analogous to steps $3$ and $4$ for the first layer). Entries in the bias vector $\rho$ corresponding to each coefficient $s_i$ were set to the expectation of the coefficient across the entire training set: $\rho_i = \mathbb{E}_t[s_{i,t}]$ (i.e., the estimate of the marginal scale parameter $\lambda_i$ for the corresponding STK). Learning was again terminated after $200000$ iterations.  To validate the learning algorithm, we ran it on a toy data set generated using known second-layer units. The algorithm successfully recovered the parameters of the generating model, as desired (see Appendix C for these results).

\subsection*{Training data and spectrogram parameters}

We trained the model on two different sound corpora. The first corpus was the TIMIT speech database \cite{timit}. The second corpus combined a set of environmental sounds (the Pitt sound database \cite{Lewicki}) and a number of animal vocalizations downloaded from freesound.org. The environmental sounds included both transient (breaking twigs, steps, etc.) and ambient (flowing water, wind, etc.) sounds; the animal vocalizations were mostly harmonic.

We computed cochleagrams by filtering sounds with a set of $65$ bandpass filters intended to mimic cochlear frequency analysis. Filters were equally spaced on an equivalent rectangular bandwidth (ERB) scale \cite{GlasbergMoore}), with parameters similar to that from a previous publication \cite{McDermottSimoncelli}. Center frequencies ranged from $200$ Hz to $8$ kHz. We computed the Hilbert envelope of the output of each filter and raised it to the power of 0.3, emulating cochlear amplitude compression\cite{Ruggero}. To reduce dimensionality, each envelope was downsampled to $320$ Hz.

We set the number of features in the first layer to $128$ and in the second layer to $100$. Pilot experiments yielded qualitatively similar results for alternative feature dimensionalities. Each first layer feature encoded a $203$ ms interval ($65$ time samples of the spectrogram). 

\section*{Results}

\subsection*{First layer: Basic spectrotemporal features of natural sounds}

\begin{figure}[H]
  \centering
  \includegraphics[scale=1]{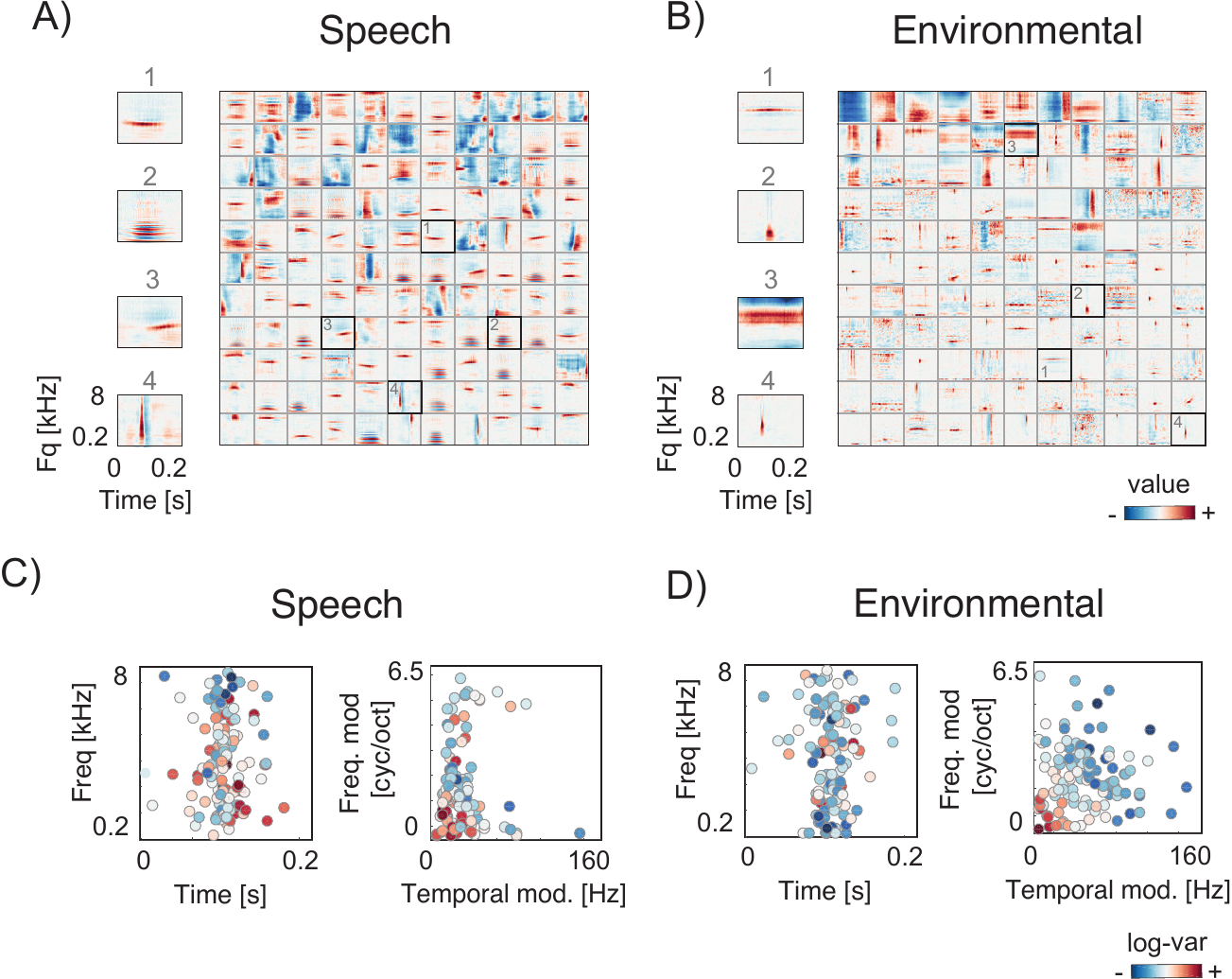}
  \caption{{\bf Spectrotemporal kernels learned by the first layer.} A) A population of STKs learned from a speech database. Representative STKs for each corpus are magnified and numbered from 1-4 for ease of reference. B) Population of STKs learned from environmental sounds. C) Speech-trained STK population plotted on time-frequency (left) and spectral-temporal modulation (right) planes. Each dot corresponds to a single STK, and its color encodes the log of its mean coefficient (averaged over entire training dataset). D) The STK population trained on environmental sounds, represented as in (C).}
\label{fig6STRFs} 
\end{figure}

The first-layer features learned from each of the two sound corpora are shown in Fig \ref{fig6STRFs} A and B. These features could be considered as the model analogues of  neural STRFs. The vast majority of features are well localized within the time-frequency plane, encoding relatively brief acoustic events. The STKs learned from speech included single harmonics and harmonic "stacks" (Fig \ref{fig6STRFs} A - features numbered 1 and 2), frequency sweeps (feature 3), and broadband clicks (feature 4). The features learned from environmental sounds also included single harmonics and clicks (Fig \ref{fig6STRFs} B - features 1, 2 and 4). In contrast to the results obtained with speech, however, harmonic stacks were absent, and a number of high-frequency hisses and noise-like features were present instead (feature 3).

The properties of the learned dictionaries are also reflected in distributions of feature locations in the time-frequency and spectrotemporal modulation planes, as visible in Fig \ref{fig6STRFs}C and D. Each dot position denotes the center of mass of a single STK, while its color signals the feature's average coefficient value over the stimulus set. For both speech and environmental sounds, the learned STKs uniformly the audio frequency spectrum (Fig \ref{fig6STRFs} C and D, left panels). Due to the convolutional nature of the code, the energy of each feature is concentrated near the middle of the time axis. The modulation spectra of the learned STKs (Fig \ref{fig6STRFs} C and D, right panels) are somewhat specific to the sound-corpus. Features trained on a speech corpus were more strongly modulated in frequency, while environmental sounds yielded STKs with faster temporal modulations. STKs learned from both datasets exhibit a spectrotemporal modulation tradeoff: if a STK is strongly temporally modulated its spectral modulation tends to be weaker. This tradeoff is an inevitable consequence of time-frequency conjugacy \cite{SinghTheunissen}, and is also found in the STRFs of the mammalian and avian auditory systems \cite{Miller,WoolleyTheunissen}.  

\subsection*{Second layer: Combinations of spectrotemporal features}

STKs captured by the first layer of the model reflect elementary features of natural sounds. By contrast, the features learned by the second layer capture how activations of different STKs cluster together in natural sounds, and thus reflect more complex acoustic regularities. We first present several ways of visualizing the multi-dimensional nature of the second-layer representation, then make some connections to existing neurophysiological data, and then derive some neurophysiological predictions from the model.

\subsubsection*{Visualizing second-layer features}
\begin{figure}[H]
  \centering
  \includegraphics[scale=1]{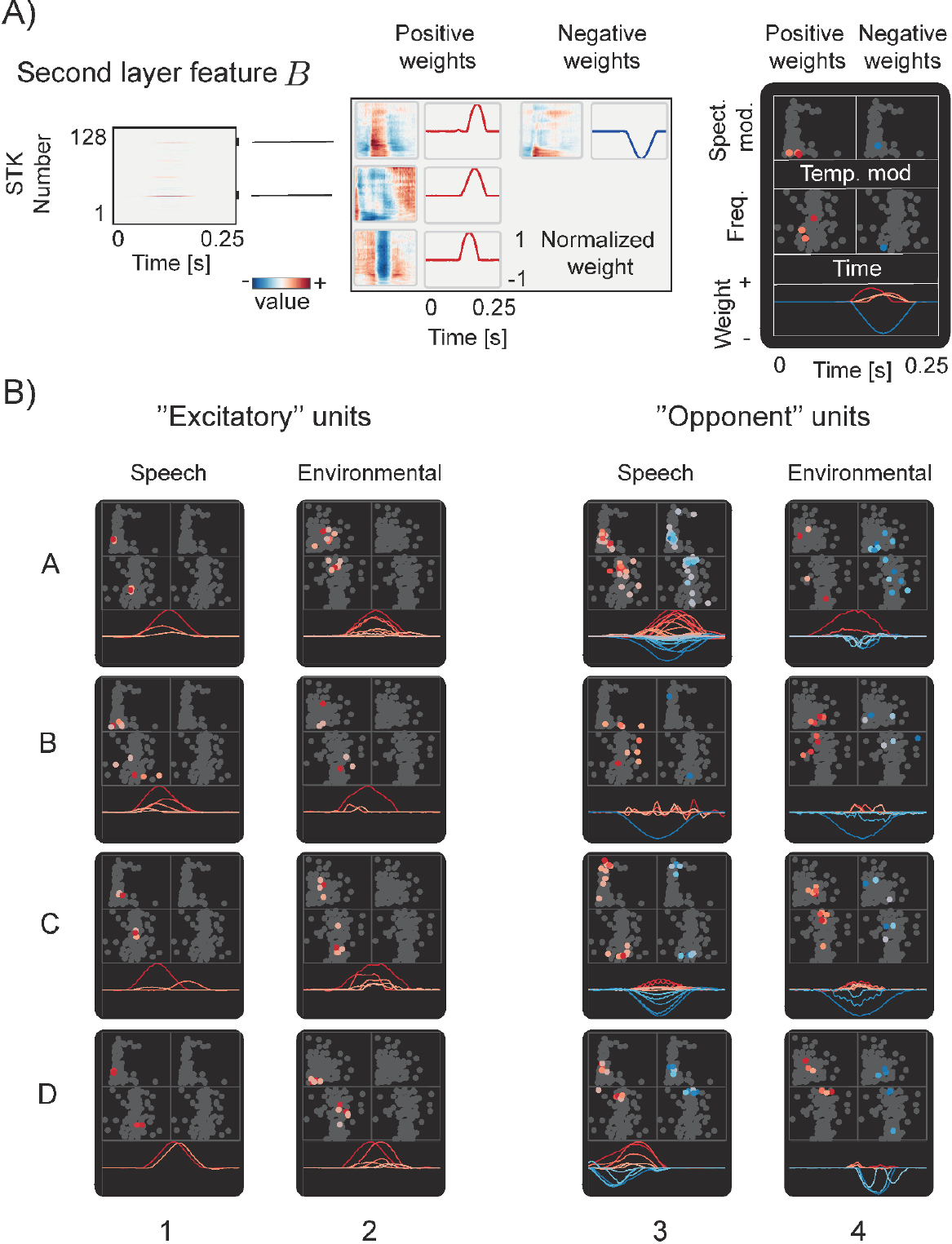}
  \caption{{\bf Second layer model features.} A) Feature visualizations. Left panel - the first layer STK weights for an example second layer feature (representing magnitudes of each STK over a time window). Middle panel - STKs whose weights in the same second-layer feature deviate most strongly from $0$ are displayed along with their weight profile over time. Right panel - STKs are plotted as dots in the modulation and time-frequency planes, with the dot location indicating the STK center of mass in the plane, and the color indicating the weight sign and magnitude (red denoting positive and blue denoting negative). STKs are divided into those with positive and negative weights for clarity. Bottom row visualizes temporal trajectories of STK weights for the feature. B) Examples of learned second-layer features. First two columns (labeled $1$ and $2$) depict units with positive ("excitatory") weights only. Last two columns (labeled $3$ and $4$) depict "opponent" units that pool features with both positive and negative weights.}
\label{fig7SecondLayerFeatures} 
\end{figure}

The second-layer features encode temporal combinations of STK log-magnitudes. Each feature can be represented as a $N \times T_b$ dimensional matrix, where rows correspond to STKs and columns to time-points. An example second-layer unit is shown in Fig \ref{fig7SecondLayerFeatures}A (left panel). A positive value in the $i$-th row and $t$-th column of a feature $B_j$ encodes a local increase in the magnitude of the $i$-th STK. A negative value encodes a decrease in magnitude. 

An alternative visualization is to examine the spectrotemporal structure of the STKs that have large weights in the second-layer feature. The same feature $B_j$ is depicted in this way in the center panel of Fig \ref{fig7SecondLayerFeatures}A, which displays the four STKs with highest average absolute weights for this particular feature. To the right of each STK are its weights (i.e., the corresponding row of the $B_j$ matrix). It is apparent that the weights increase and decrease in a coordinated fashion, and thus likely encode particular dependencies between the STKs.

To summarize the full distribution of STKs contributing to a second-layer unit, we adopted the visualization scheme illustrated in the right panel of Fig \ref{fig7SecondLayerFeatures}A. We plot the center of mass of each STK in the modulation (top row) and time-frequency (middle row) planes, as in Fig \ref{fig6STRFs}C and D. The dot for a first-layer STK is colored red or blue, depending on the sign of their time-averaged weight, with the average absolute value of the weight signaled by the intensity of the color. The bottom row of the panel depicts the temporal pattern of STK magnitudes - line colors correspond to dots in the top and middle rows of the panel. Although the weights of most STKs maintain the same sign over the temporal support of the second-layer unit, they were not directly constrained in this regard, and in some cases the weight trajectories cross zero.

Representative examples of second-layer basis functions are depicted in this way in Fig \ref{fig7SecondLayerFeatures}B. We separated them into two broad classes - "excitatory" units (columns $1$ and $2$ in Fig \ref{fig7SecondLayerFeatures}), which pool STKs using weights of the same sign, and therefore encode a pattern of coordinated increase in their magnitudes, and "excitatory-inhibitory" units (columns $3$ and $4$ in Fig \ref{fig7SecondLayerFeatures}) which pool some STKs with positive average weights and others with negative average weights. We note that excitatory-only and inhibitory-only units are functionally interchangeable in the model, because the encoding is unaffected if the sign of both the STK weights and coefficients are reversed. From inspection of Fig \ref{fig7SecondLayerFeatures}B it is evident that some second-layer units pool only a few STKs (e.g. A1, C1, D1) while others are more global and influence activations of many first-layer units (e.g. A2, A3, C3).

\subsubsection*{Second layer features encode patterns of STK dependencies}

To better understand the structure captured by the second-layer units, we considered their relationship to the two kinds of dependencies in STK activations that initially motivated the model. In Methods and Models section we observed that  STK activations to natural sounds exhibit strong local cross-correlations as well as variations of particular temporal activation patterns. The second layer of the model was formulated in order to capture and encode these redundancies.

\begin{figure}[H]
  \centering
  \includegraphics[scale=1]{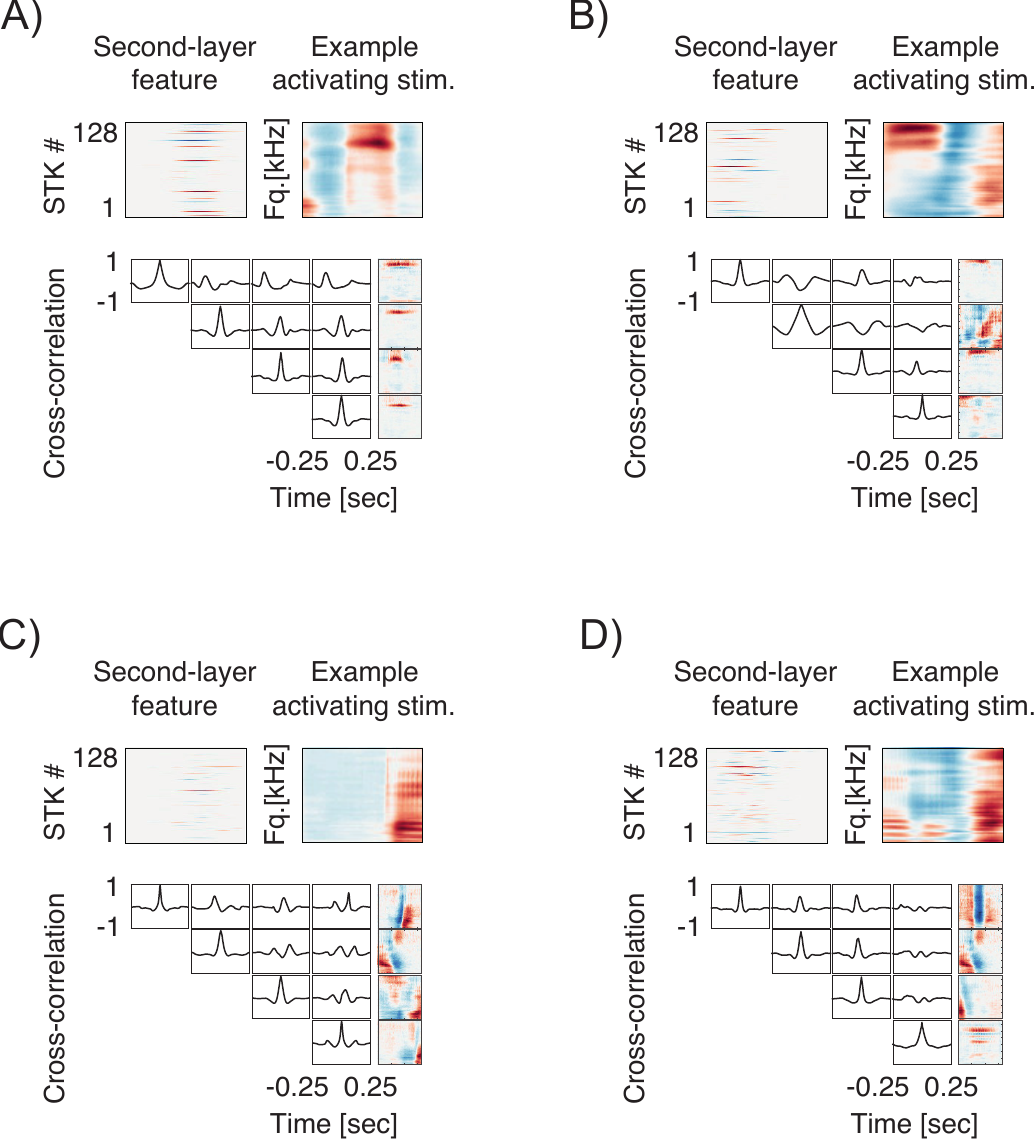}
  \caption{{\bf Second-layer units respond to specific STK cross-correlation patterns.}A) A second-layer unit (top row, left column) and an example sound excerpt eliciting a strong positive response in the unit (top row, right column).  Bottom shows cross-correlation functions of coefficient trajectories for four STKs. The STKs selected were those with the largest weights for this second-layer unit. Cross-correlations were averaged across $25$ stimuli eliciting the strongest positive response of the second-layer unit across a large subset of the TIMIT corpus.  B, C, D) same as A for three other second-layer units.}
\label{fig10CrossCorr} 
\end{figure}

To first test whether the second layer captures the sorts of residual correlations evident in the first layer output, we measured cross-correlations between STK activations conditioned on the activation of particular second-layer units. Fig. \ref{fig10CrossCorr}A-D depicts four example second-layer units (top row, left column) along with an example stimulus that produced a strong positive response in the unit (selected from the TIMIT corpus). The bottom section of each panel depicts cross-correlation functions of activations of four STKs, averaged over $25$ stimulus epochs that produced a strong positive response of the second-layer unit. The cross-correlation functions deviate substantially from $0$, as they do when conditioned on excepts of natural sounds (Fig. \ref{fig3Dependencies}). These correlations reflect the temporal pattern of STK coefficients that the second-layer unit responds to.

\begin{figure}[H]
  \centering
  \includegraphics[scale=1]{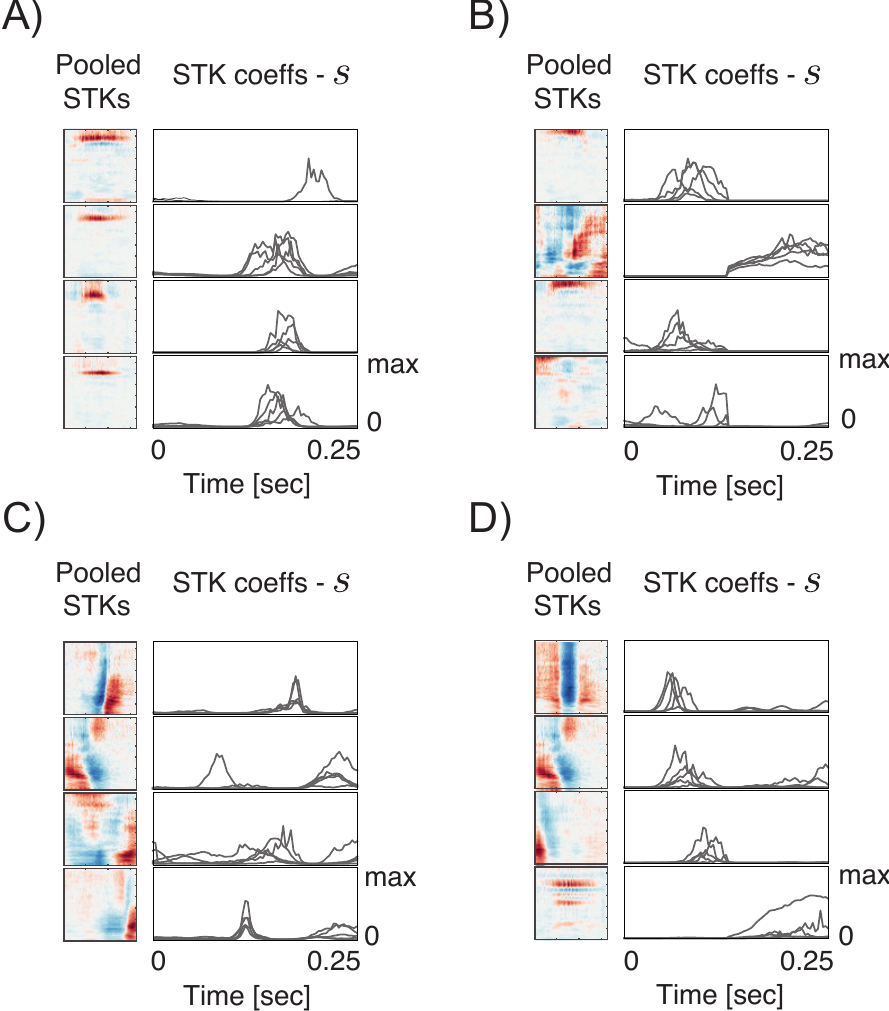}
  \caption{{\bf Second-layer units generalize across variation in STK coefficients.}Panels A-D correspond to second-layer units depicted in Fig. 8  Each panel shows four STKs pooled with strongest weights (left column) by the corresponding second-layer unit. Next to each STK are their activation patterns (right column) for the $5$ stimuli eliciting the strongest response in the second-layer unit. Despite some variability they share a global trend. }
\label{figAmplitudes} 
\end{figure}

We next examined whether the second-layer units respond to STK activations fluctuating around particular global patterns  (as depicted in Fig.\ref{fig4Variance}). Because the second layer of the model represents the magnitude, rather than the precise values of first-layer coefficients, it should be capable of generalizing over minor STK coefficient variation. Fig.\ref{figAmplitudes} plots STK coefficient trajectories for stimuli eliciting a strong response in the second-layer features from Fig. \ref{fig10CrossCorr}. The STK activation traces reveal variability in each case, but nonetheless exhibit a degree of global consistency, as we saw earlier for natural sound exemplars (Fig.\ref{fig4Variance}). These results provide evidence that the second layer is capturing the dependencies it was intended to model. 

\subsection*{Experimental predictions}

\subsubsection*{Pooling of similiar spectrotemporal patterns}

What do the model results suggest about mid-level auditory structure? Our visualizations of second layer features in Fig \ref{fig7SecondLayerFeatures}B revealed that many represent the concurrent activation of many STKs. Examination of Fig \ref{fig7SecondLayerFeatures}B (columns 1 and 2) suggests that the first layer STKs that such units pool are typically similar either in their spectrotemporal or modulation properties. For example, the unit depicted in panel A2 encodes joint increases in the magnitude of high-frequency STKs, while that in panel C2 encodes joint increases in low-frequency STKs. 

In order to quantitatively substantiate these observations, we measured the spread of the center-of-mass of each of the STKs pooled by "excitatory-only" second-layer units (those whose STK weights exceeding $5\%$ of the maximum absolute weight were all of the same sign). Specifically, we computed the standard deviation of the center-of-mass on each dimension of the time-frequency and modulation planes, for all STKs with weights exceeding $5\%$ of the maximum weight for the unit, and compared this to the standard deviation of centers of mass for random samples of STKs (matched in size to the mean number of STKs pooled by the excitatory-only units). These standard deviations measure the spread of STKs in spectrotemporal and modulation domains; if the pooled STKs are similar, the spread should be small. 

As shown in Fig.\ref{figPoolingStats}, the STKs pooled with the same sign were more similar in each of the four analyzed dimensions than random groups of STKs, for both sound corpora (red bar vs. gray bar, t-tests yielded p<.05 in all cases). This result suggests the hypothesis that STRFs that are pooled by downstream auditory cortical neurons with excitatory weights \cite{AtencioExpands,KozlovGentnerComposite} should also tend to be more similar than expected by chance.

\subsubsection*{Opponency patterns in mid-level audition}

Fig \ref{fig7SecondLayerFeatures}B also shows examples of a distinct set of second-layer units in which increased activation of one group of STKs (again, typically similar to each other) is associated with a decrease in activity of another group of STKs. For example, the unit in panel B3 encodes an increase in magnitude of temporally modulated features (clicks) together with a simultaneous decrease of activation of a strongly spectrally modulated (harmonic) STK. Such "opponency" was evident in a large subset of second layer units, and represents the main novel phenomenon evident in our model. 

To our knowledge no such opponent tuning has been identified in auditory neuroscience, but qualitatively similar opponent behavior is evident in visual neurons exhibiting end-stopping or cross-orientation inhibition. The results raise the possibility that coordinated excitation and inhibition could be a feature of central auditory processing, and we thus examined this model property in detail.

To quantitatively substantiate the observation of opponency, we again examined the variation in STKs pooled by the second-layer units, in this case those that pooled STKs with both positive and negative weights. We analyzed all units that had at least one STK weight of each sign whose absolute value exceeded $5\%$ of the maximum weight. For each such second-layer unit we identified centers of mass of the STKs pooled with weights exceeding $5\%$ of the maximum and again computed their standard deviation along time-frequency and modulation coordinates. We compared the standard deviation for STKs pooled with the same sign (separately computed for positive and negative signs) to that of STKs pooled with both signs. If opponent units contrast different types of features we expect the standard deviation of STKs pooled with both signs to be higher than that of STKs pooled with a single sign. 

\begin{figure}[H]
  \centering
  \includegraphics[scale=1]{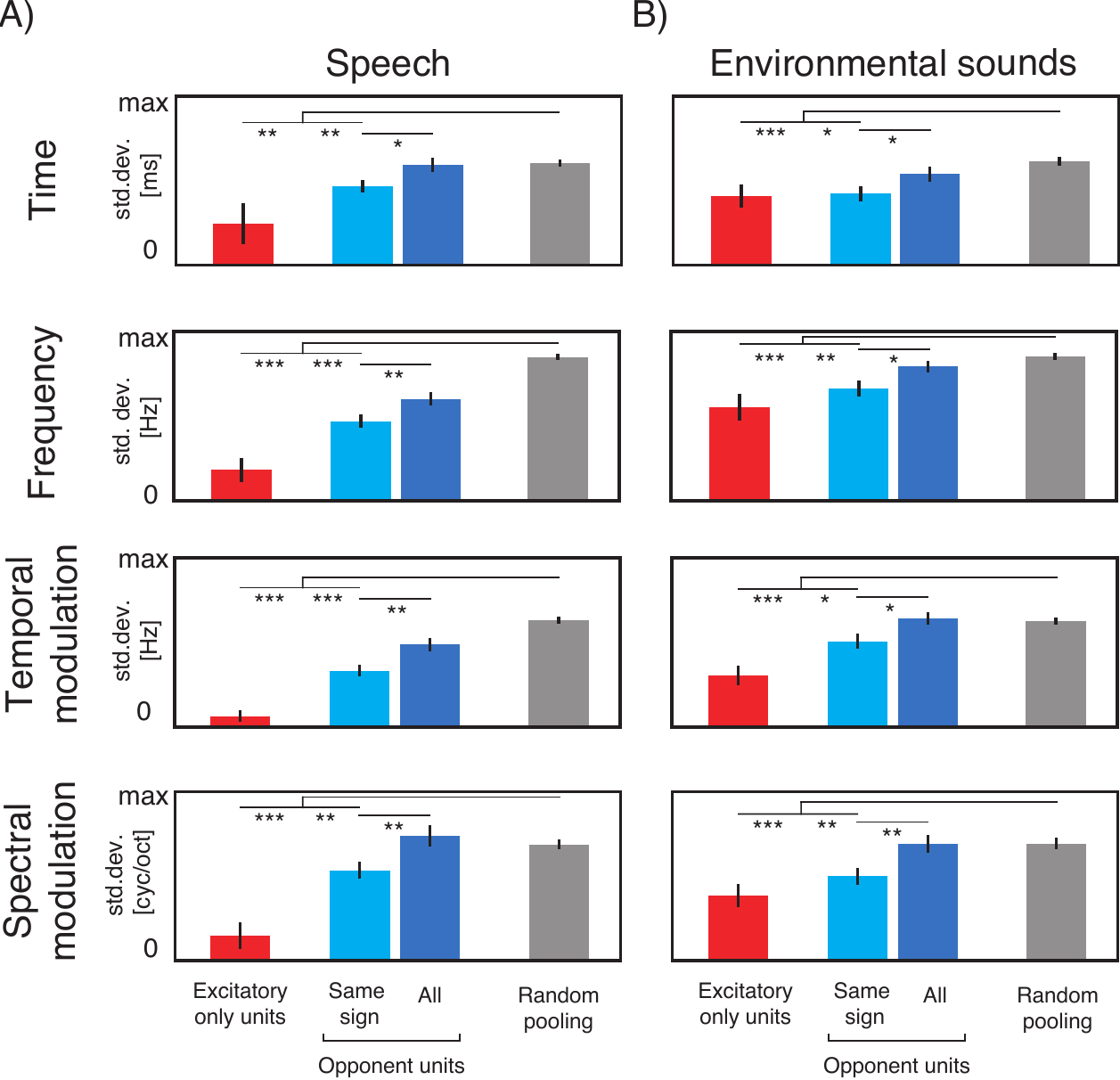}
  \caption{{\bf Statistics of STK pooling by second layer units.}
Average standard deviation (spread) of centers of mass of STKs pooled by second layer units. Red bars plot spread of STKs pooled by excitatory-only units. Blue bars plot spread of STKs pooled by opponent units. Light and dark blue bars plot the spread of STKs pooled by opponent units with the same sign (light blue) and with either sign (dark blue). Gray bars plot the spread of random subsets of STKs equal in size to the average number of STKs pooled by second layer units. Error bars plot the standard error of the mean. A) Pooling statistics for model trained on speech. B) Pooling statistics for model trained on environmental sounds.}
\label{figPoolingStats} 
\end{figure}

As shown in Fig.\ref{figPoolingStats}, even for STKs pooled with the same sign, the spread was typically higher for opponent units than for excitatory-only units, for both speech and environmental sounds (blue vs. red bars in Fig.\ref{figPoolingStats}A and B respectively). However, the average spread of STKs pooled with the both signs by opponent units (Fig.\ref{figPoolingStats}, dark blue bars) was significantly higher than the spread of STKs pooled with the same sign (Fig.\ref{figPoolingStats}, light blue bars). This finding held for all dimensions and both corpora ($p<.05$ in all cases, via t-test). ). Moreover, the  spread of STKs pooled with the same sign by opponent units were also significantly lower than the spread of random groups of STKs (light blue bars vs. gray bars; $p<.05$ in all cases, via t-test). These results provide quantitative evidence that second-layer units pool features in a structured way and that opponent units assign opposite signs to to groups of STKs that are similar within groups but differ across groups.

\begin{figure}[H]
  \centering
  \includegraphics[scale=.6]{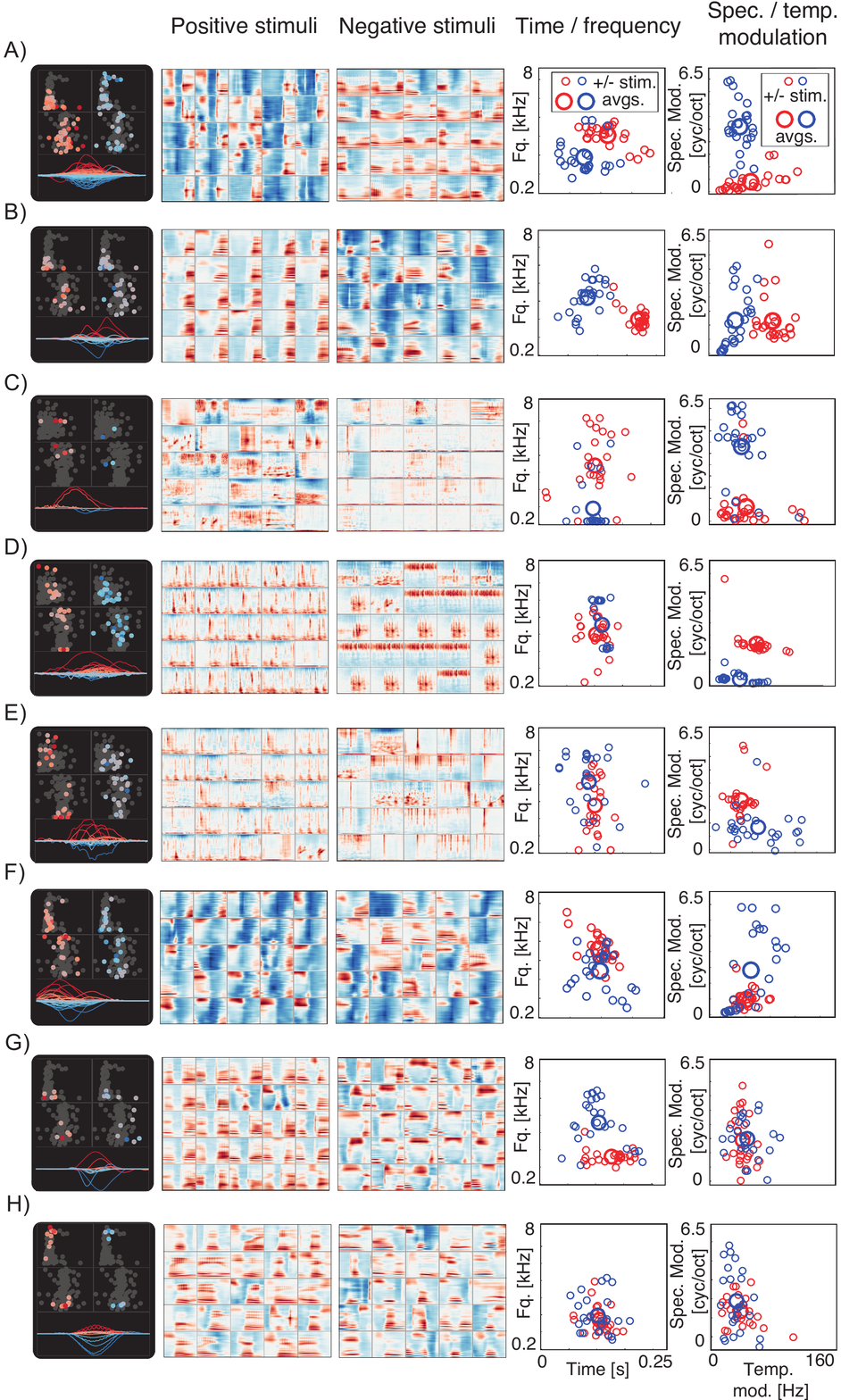}
  \caption{{\bf Opponent stimulus selectivity in second layer units.} Each row corresponds to a particular second-layer unit. The leftmost column plots the center of mass of each first-layer STK in the modulation and time-frequency planes, along with the time courses of their weights in the second-layer unit (as in Fig \ref{fig7SecondLayerFeatures}B). The second and third columns from the left depict $25$ stimuli eliciting strong positive and negative responses in the corresponding second-layer unit. In the fourth and fifth columns, positive and negative stimuli are visualized as red and blue circles, respectively, in time-frequency and modulation planes (the circle is located at the center of mass of the stimulus). Large circles correspond to centroids of positive and negative stimulus clusters.}
\label{fig9Opposing} 
\end{figure}

One might imagine that examination of the STKs pooled by each unit would be sufficient to determine the sort of stimuli eliciting strong positive or negative responses. But because the activation of each unit is the result of a non-linear inference process in which units compete to explain the stimulus pattern \cite{OlshausenField}, it is often not obvious what a set of STKs will capture. Thus to understand which stimuli 'excite' or 'inhibit' second-level features, we inferred coefficients $v$ by encoding the entire training dataset and selected two sets of $25$ sound epochs that elicited the strongest positive and strongest negative responses, respectively, in each unit. Examples of opponent stimulus patterns encoded by second-layer features are visualized in Fig \ref{fig9Opposing}. These positive and negative stimuli are depicted in the second and third columns from the left, respectively. In the last two columns, the center of mass of each of these stimuli is plotted on the time-frequency plane (fourth column) and modulation plane (fifth column). Although the center of mass of each stimulus is admittedly a crude summary, the simplicity of the representation facilitates visualization and analysis of the clustering of positive and negative stimuli. We note also that because second-layer coefficients and weights can be sign-reversed without changing the representation, the designation of stimuli (and STK weights) as 'excitatory' or 'inhibitory' is arbitrary.

In some cases, some natural function can be ascribed to the unit, particularly upon listening to stimuli eliciting positive and negative responses. For instance, positive activations of the unit shown in Fig \ref{fig9Opposing}B appear to encode onsets of voiced speech, while its negative activations encode voicing offsets. By contrast, the unit shown in Fig \ref{fig9Opposing}H appears to code speaker gender. For this unit, representations of positive and negative stimuli were mixed on the time-frequency and modulation planes, but the excitatory stimuli exhibited somewhat coarser spectral modulation (red circles in the right column are more concentrated in the lower part of the plane than blue circles). Listening to the stimuli revealed a clear difference in pitch/gender (23 out of 25 of the negative stimuli were vowels uttered by a female speaker, while all of the positive stimuli were vowels uttered by a male speaker, as confirmed by the first author upon listening to the stimuli). This observation underscores the fact that the time-frequency and modulation planes are but two representational spaces in which to assess stimuli, and are used here primarily because they are standard. Moreover, the centroids in these planes are but one way to summarize the STK. Although opponent sets of STKs and positive/negative stimuli often exhibited separation in one of the two planes when analyzed in terms of centroids, they need not do so in order to be distinct.

Taken together, our results demonstrate opponent units that encode STK activation patterns which are mutually exclusive and presumably do not co-occur in natural signals. The phenomenon is a natural one to investigate in the auditory cortex.

\subsection*{Comparison with neurophysiological data}

Although our primary goal was to generate predictions of not-yet observed neural representations of sound, we also sought to test whether our model would reproduce known findings from auditory neuroscience. The first layer STKs replicated some fairly standard findings in the STRF literature, as discussed earlier. To compare the results from the second layer to experimental data, we examined their receptive field structure and the specificity of their responses, and compared each to published neurophysiology data. 

\begin{figure}[H]
  \centering
  \includegraphics[scale=1]{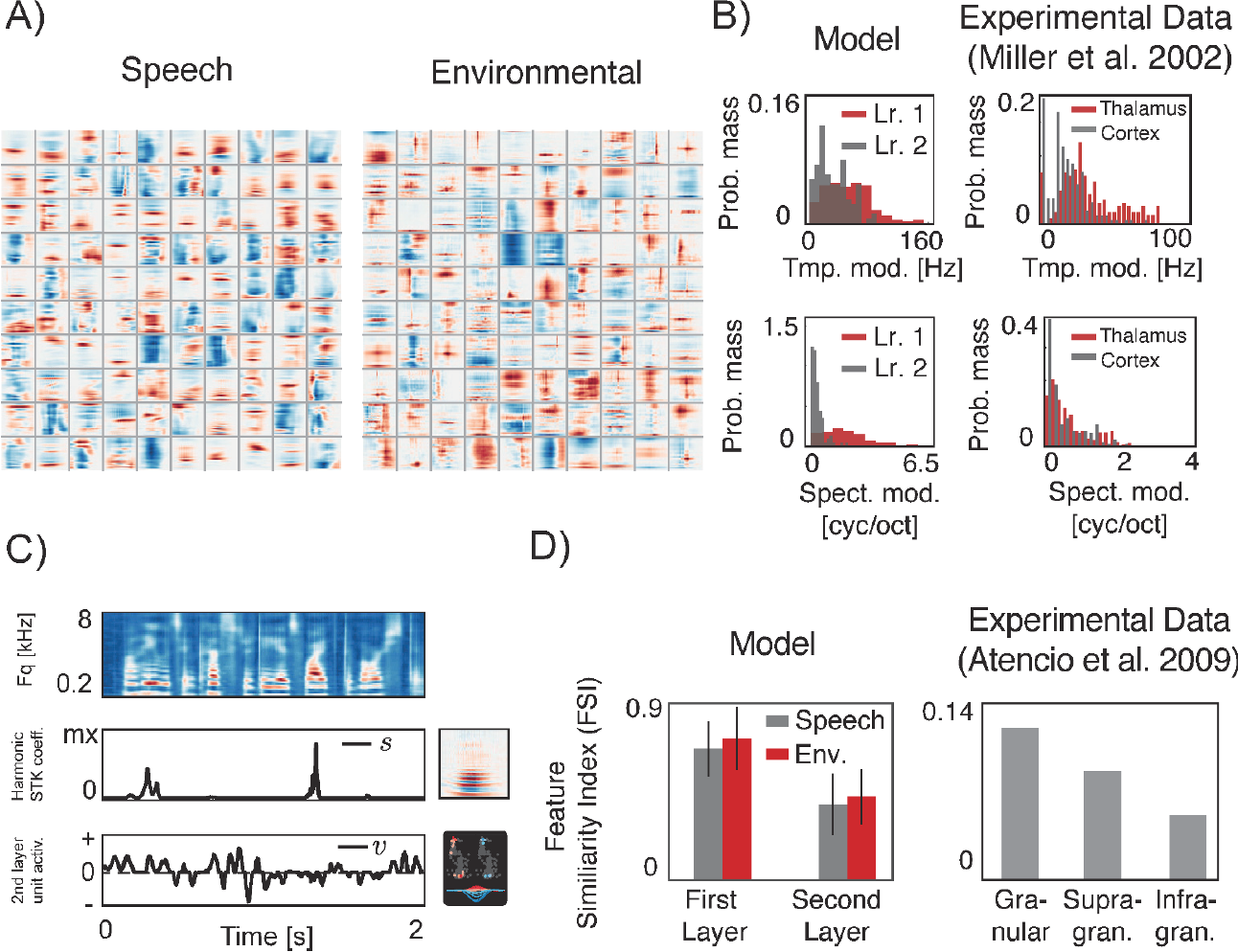}
  \vspace{0.05cm}
  \caption{{\bf Comparison with neurophysiological data.} A) Spectrotemporal receptive fields estimated for second layer units trained on speech and environmental sounds. B) Comparison of spectral and temporal modulation tuning of first and second layer receptive fields (panels in left column) to experimental measurements in auditory thalamus and cortex of the cat (panels in right column were provided by Lee Miller \cite{Miller}). Higher processing stages both in the model and the auditory system exhibit tuning for coarser modulations in both frequency and time. Sub-panels with experimental data reprinted with permission of original author. C) Example activation trajectories of first (middle row) and second layer (bottom row) units to an excerpt of speech (top row). The activation of the first layer unit is tightly locked to presence of a preferred stimulus. Responses of the second layer unit are less specifically locked to particular spectrotemporal structures. D) Comparison of tuning specificity, as measured with the Feature Selectivity Index (FSI), for first and second layers of the model trained either on speech or environmental sounds (gray and red bars, respectively), and for different layers of the auditory cortex of the cat (right panel replotted from \cite{Atencio}).}
\label{fig8Comparison} 
\end{figure}
\pagebreak

\subsubsection*{Differences in modulation tuning between first and second model layers}
First, we estimated spectrotemporal receptive fields for units in both layers of the model. The receptive fields of the first-layer units are simply the STK of the unit. To estimate receptive fields of a second-layer unit, we drew inspiration from the spike-triggered average, generating a number of cochleagram samples from each basis function and averaging them. To generate samples from the $j-$th basis function, we set a coefficient $v_j$ to $1$ with all other $v_{i \neq j}$ set to zero. We then sampled STK activation trajectories from the distribution dictated by the second-layer feature's coefficient and weights, convolved them with the corresponding STKs and summed the results. We then averaged multiple such samples together. Although we could have computed something more directly analogous to a spike-triggered average, the average sample (which we can compute only because we have the underlying generative model, unlike when conducting a neurophysiology experiment) has the advantage of alleviating the influence of stimulus correlations on the signature of the receptive field. See Appendix B \ref{AppendixB} for an illustration of this analysis.

"Receptive fields" obtained in this way are depicted in Fig \ref{fig8Comparison}A. To compare the model units to neurophysiological data, we generated histograms of average spectral and temporal modulation frequency (center of mass in the modulation plane) of first- and second-layer receptive fields and plot them next to distributions of preferred modulation frequencies of neurons in the auditory thalamus and cortex of the cat \cite{Miller} (Fig \ref{fig8Comparison}B). The same trend is evident in the model and the auditory system: the second-layer prefers features with slower/coarser spectral and temporal modulations relative  to the first-layer, mirroring the difference seen between the cortex and thalamus. This analysis used features trained on speech, but environmental sounds yielded qualitatively similar results. In the model as well as the brain, lower modulation frequencies may result in part from combining multiple distinct STKs in downstream units. 

\subsubsection*{Model units reflect neural hierarchical trends of response specificity} 
Neuronal tuning in early and late stages of the auditory system also tends to differ in specificity \cite{Chechik, KozlovGentnerComposite}. Compared to the auditory brainstem, cortical neurons are less selective and respond to multiple features of sound \cite{KozlovGentnerComposite,AtencioExpands}, consistent with an increase in abstraction of the representation \cite{Chechik}. Suggestions of similar behavior in our model are apparent in the activations of first- and second-layer units to sound, an example of which is shown in Fig \ref{fig8Comparison}C. The first layer feature (middle row) becomes activated only when it is strongly correlated with the stimulus. In contrast, activations of a typical second layer feature (bottom row) deviate from zero during many, seemingly different parts of the stimulus.

We quantified the specificity of tuning with the feature selectivity index (FSI), a measure introduced previously to quantify how correlated a stimulus has to be with a neuron's spike-triggered average to evoke a response \cite{MillerFSI} (see Appendix B \ref{AppendixB} for definition). An FSI equal to $1$ implies that a neuron spikes only when a stimulus is precisely aligned with its STRF (defined as the spike-triggered average), whereas an FSI equal to $0$ means that the stimuli triggering neural firing are uncorrelated with the STRF. We computed the FSI using the $25$ cochleagram excerpts that most strongly activated each of the first- and second-layer units. The average FSI of each model layer is plotted in Fig \ref{fig8Comparison}D. Second-layer features are substantially less specific than first-layer features. A similar effect occurs across different cortical layers \cite{Atencio} (Fig \ref{fig8Comparison}D, right). Analogous differences seem likely to occur between thalamus and cortex as well, although we are not aware of an explicit prior comparison. The decrease in response specificity in our model can be explained by the fact that second layer units can become activated when any of the pooled first-layer features (or their combination) appears in the stimulus.

\section*{Discussion}

Natural sounds are highly structured. The details of this structure and the mechanisms by which it is encoded by the nervous system remain poorly understood. Progress on both fronts is arguably limited by the shortage of signal models capable of explicitly representing natural acoustic structure. We have proposed a novel statistical model that captures an unexplored type of high-order dependency in natural sounds – correlations between the activations of basic spectrotemporal features. Our model consists of two layers. The first layer learns a set of elementary spectrotemporal kernels and uses them to encode sound cochleagrams. The second layer learns a representation of co-occurence patterns of the first layer features. 

We adopted a generative modelling approach, inspired by its previous successes in the domain of natural image statistics. Previous hierarchical, probabilistic models of natural images were able to learn high-order statistical regularities in natural images \cite{Karklin, KarklinVariance, Olshausen, Hyvarinen, HyvarinenHoyerContours, DeepBeliefV2, HyvarinenHosoya, GarriguesOlshausen,BerkesTurnerSahani}. In many cases the representations learned by these models exhibit similarity to empirically observed neural codes in the visual system.

The hierarchical representations learned by our model provide predictions about neural representations of mid-level sound structure. Moreover, the model reproduces certain aspects of the representational transformations found through the thalamus and cortex. The results suggest that principles of efficient coding could shape mid-level processing in the auditory cortex in addition to the auditory periphery \cite{Lewicki}. 

\subsection*{Statistical dependencies in natural sounds}

Our model exploits statistical dependencies between first-layer spectrotemporal kernels. These dependencies have two likely causes. First, dependencies almost surely remain from limitations of the convolutional sparse coding model, in that first-layer coefficients are never completely marginally independent even after learning (due to insufficient expressive power of the code). Second, even if the first-layer coefficients were fully marginally independent they could exhibit local dependencies when conditioned on particular sound excerpts. Observations of such conditional dependence have been made previously, mostly in the context of modelling natural image statistics \cite{Karklin, KarklinVariance,Hyvarinen,Olshausen,KarklinEkanadham}. We similarly observed strong non-zero cross-correlations between STKs for particular natural sound excerpts, and found that the second-layer units captured these sorts of dependencies (Fig. \ref{fig3Dependencies},\ref{fig10CrossCorr}).

We also observed that different instances of the same acoustic event (such as the same word uttered twice by the same speaker) yield similiar STK coefficient trajectories (Fig. \ref{fig4Variance}), which can be thought of as samples from a nonstationary distribution with a time-varying magnitude parameter. By modeling this magnitude the model learned representations with some degree of invariance (Fig. \ref{figAmplitudes}).

\subsection*{Model results - Convolutional spectrotemporal features}

The model first learned a sparse convolutional decomposition of cochleagrams. Convolutional representations are less redundant than “patch-based” codes and can represent signals of arbitrary length, and have been succesfully applied to spectrogram analysis in engineering (e.g. \cite{ConvSparse, ConvSparse2}). The features learned with convolutional sparse coding improved sound classification accuracy, musical note extraction and source separation in prior work \cite{ConvSparse,ConvSparse2}. By contrast, the model presented here is, to our knowledge, the first use of convolutional sparse coding of audio in a neuroscience context. Accordingly, we provide an in-depth analysis of feature shapes learned by the first layer and relate them to known spectrotemporal tuning properties of auditory neurons. 

The learned spectrotemporal kernels span clicks, harmonics, combinations of harmonics, bandpass noise, frequency sweeps, onsets, and offsets. Some of these structures are present in previous learned codes of spectrograms, but due to the convolutional nature of the learned code, the model learns only a single version of each feature rather than replicating it at different time points. Although difficult to quantify, our impression is that the code is more diverse than that obtained with previous patch-based approaches \cite{Carlson,klein2003sparse}. As in previous neurophysiological measurements of cortical STRFs, the model STKs tile the spectrotemporal modulation plane, subject to the contraints of the tradeoff between time and frequency \cite{SinghTheunissen}.

As with features learned from sound waveforms \cite{Lewicki,SmithLewicki}, we found the first-layer spectrogram features to depend on the training corpus (speech or a set of environmental sounds). Although certain spectrotemporal patterns (e.g. single harmonics or clicks) appeared in both features sets, others were corpus-specific. Corpus-specific structure was also evident in the distributions of features in the modulation plane. This corpus dependence contrasts with the relatively consistent occurrence of Gabor-like features in sparse codes of images. These results raise the possibility that auditory cortical neuronal tuning might exhibit considerable heterogeneity, given that the full range of natural audio must be encoded by cortical neurons.

\subsection*{Model results – Second-layer features}

The second layer of the model encodes a probability distribution of spectrotemporal feature activations. By jointly modelling the magnitudes of the first layer responses, the second layer basis functions capture patterns of spectrotemporal kernel covariation. Magnitude modeling was essential to learning additional structure – we found in pilot experiments that simply applying a second layer of convolutional sparse coding did not produce comparable results. This observation is consistent with previous findings that nonlinear transformations of sparse codes often help to learn residual dependencies \cite{Karklin,RecursiveICA, Hyvarinen}. 

The second layer of our model learned a representation of spectrotemporal feature co-activations that frequently occur in natural sounds. Typically, the spectrotemporal features pooled by a second-layer unit shared some property, often evident in at least one of the time-frequency or modulation planes - e.g. frequency content, temporal pattern or modulation characteristics (Fig. \ref{fig7SecondLayerFeatures}), quantified by the similarity in these planes. Our model also identified 'opponent' patterns – sets of spectrotemporal features that are pooled with opposite sign, presumably because they are rarely active simultaneously in natural sounds (Fig. \ref{fig9Opposing}). The opponent features for a second-layer unit were often at least partially separated in at least one of the spectrotemporal modulation or time-frequency planes (\ref{figPoolingStats}). 

\subsection*{Relation to auditory neuroscience}

The structure learned by the model from natural sounds replicates some known properties of the auditory system. The modulation frequencies preferred by units dropped from the first to the second layer (Fig. \ref{fig8Comparison}B), as has been observed between the thalamus and cortex. Unit “tuning” specificity  (i.e. the similarity between stimuli eliciting a strong response) also decreased from the first to the second layer (Fig. \ref{fig8Comparison}C,D). A similar specificity decrease  has been observed between granular and supra- and infra-granular layers of the cortex \cite{Atencio}. Although we do not suggest a detailed correspondence between layers of our model and particular anatomical structures, these similarities indicate that some of the principles underlying hierarchical organization of the auditory pathways may derive from the natural sound statistics and architectural choices that constrain our model.

The combinations of spectrotemporal features learned by our model provide hypotheses for neural tuning that might be present in the auditory system. Some units combined only STKs with similar acoustic properties (Fig. \ref{fig7SecondLayerFeatures}, \ref{figPoolingStats}). Others encode activations of 'opponent' sets of first-layer kernels. Such opponent kernel sets are presumably those that do not typically become active simultaneously in the training corpus. The results are somewhat analogous to excitation-inhibition phenomena in visual neurophysiology (such as end-stopping, length and width suppression or cross-orientation inhibition) emerging in models of natural images \cite{Karklin, CoenCagli, ZhuRozell, HyvarinenHosoya}. The second-layer units of our model provide candidates of potentially analogous phenomena in the auditory cortex.

Our results are also relevant to recent evidence that central auditory neurons in some cases are driven by more than one stimulus feature \cite{Atencio,SharpeeHier,KozlovGentnerComposite}. Our model reproduces that trend, pooling up to dozens of single layer features with strong weights, and predicts that distinct dimensions may be combined in opponent fashion. 

\subsection*{Relation to other modeling approaches}

The model described here represents one of several approaches to construction of hierarchical signal representations. The representations in our model are learned from natural sounds, and thus contrast with hand-engineered models that seek to replicate known or hypothesized features of sensory coding \cite{ChiRuShamma,McDermottSimoncelli}. One advantage of learning models from natural signals is that any similarities with known neural phenomena provide candidate normative explanations for these phenomena, e.g. that they arise from the demands of efficient coding or some other optimality constraint imposed by the model \cite{OlshausenField, Lewicki}. Another potential advantage is that one might hope to learn structures that have not yet been observed neurally, but that could provide hypotheses for future experiments. The latter was the main motivation for our modeling approach.

The model is also unsupervised, and generative, specifying a joint probability distribution over the data and coefficients in both latent layers. An alternative approach to learning hierarchical representations of data is to use discriminative models, which optimize a representation for performance of a particular task. Prominent recent examples of such discriminative learning come from the field of deep neural networks \cite{Yamins}. Models of this class are comparatively straightforward to optimize and can yield high performance on classification tasks, but typically require large numbers of labeled training examples. The requirement of labeled training data is sometimes a practical limitation, and raises questions about the extent to which the learning procedure could relate to the brain. Generative models currently have the disadvantage of requiring custom optimization procedures. However, their generative nature facilitates certain applications (e.g. denoising) and allows samples to be straightforwardly generated, potentially for use in experiments. Moreover, they do not require labeled data, and learn representations that are independent of any particular task, much like the unsupervised learning believed to occur in sensory systems.

\section*{Conclusion}

We have presented a hierarchical model of natural sounds.  When trained on natural sound corpora, the model learned a representation of spectrotemporal feature combinations. The properties of the model layers resemble aspects of hierarchical transformations previously observed in the brain, suggesting that efficient coding could shape such transformations throughout the auditory system. The learned mid-level features provide hypotheses for auditory cortical tuning as well as a means to parameterize stimuli with which to probe mid-level audition.

\section*{Appendix A}
\label{AppendixA}

\paragraph*{Gradients for learning and inference in the first layer}

Learning and inference in the first layer was achieved via gradient descent on negative log-posterior (Eq \ref{eq3:E1}). Below we provide expressions of $E_1$ 
gradient with respect to $s_{i, t}$ and $\phi_{i, f, \tau}$ respectively

The non-negativity constraint on sparse coefficients $s$ complicates the optimization process in learning and inference. To alleviate this complication, we introduced auxiliary coefficients $z_i$ and assumed that non-negative sparse coefficients $s_i$ are equal to squares of $z_i$:
\begin{equation}
s_i = z_i^2, z_i \in \mathbb{R}
\end{equation}
We replace $s_i$ with $z_i^2$ in equations below.

Gradients of first layer energy function (Eq \ref{eq3:E1}) with respect to $z_{i,t}$ and $\phi_{i, f, \tau}$ was respectively:

\begin{equation}
\label{dE1dz}
\frac{\partial E_1}{\partial z_{i,t}} \propto -\frac{4 z_{i,t}}{\sigma^2} \sum_{f=1}^F (\phi_{i,f} \odot e_{f})_t + 2 \lambda_i z_{i,t}
\end{equation} 

\begin{equation}
\label{dE1dPhi}
\frac{\partial E_1}{\partial \phi_{i,f,\tau}} \propto -2 \Big(s_{i} \odot e_{f} \Big)_{\tau}
\end{equation}

where $\odot$ denotes cross-correlation and $e_{f,t} = x_{f,t} - \hat{x}_{f,t}$ is the reconstruction error.

\paragraph*{Gradients for learning and inference in the second layer}

Similarly, learning and inference in the second layer was achieved via gradient descent on the corresponding energy function (equation \ref{eq7:E2}). Gradient expressions took the following form:

\begin{equation}
\label{dE2dV}
\frac{\partial E_2}{\partial v_{i,t}} \propto \sum_{n=1}^N \psi_n \odot B_{i,n} + \alpha \sgn (v_{i,t}) 
\end{equation}

\begin{equation}
\label{dE2dB}
\frac{\partial E_2}{\partial B_{j,i,t_b}} \propto   (1 - \psi_i \odot v_j)_{t_b} + \beta \sgn(B_{j,i,t_b})
\end{equation}

where $\hat{\lambda}_{n,t} = \lambda_n \exp\left[\sum_j^M v_j \ast B_j \right]_t$ is the reconstruction of instantaneous scale parameter and $\psi_{n,t} = \frac{s_{n,t}}{\hat{\lambda}_{n,t}}$.
\pagebreak

\section*{Appendix B}
\label{AppendixB}

In this appendix we describe the methods used to estimate and analyze receptive fields of model units.

\subsubsection*{Receptive field estimation}
\renewcommand{\thefigure}{B\arabic{figure}}
\setcounter{figure}{0}

\begin{figure}[H]
  \centering
  \includegraphics[scale=.7]{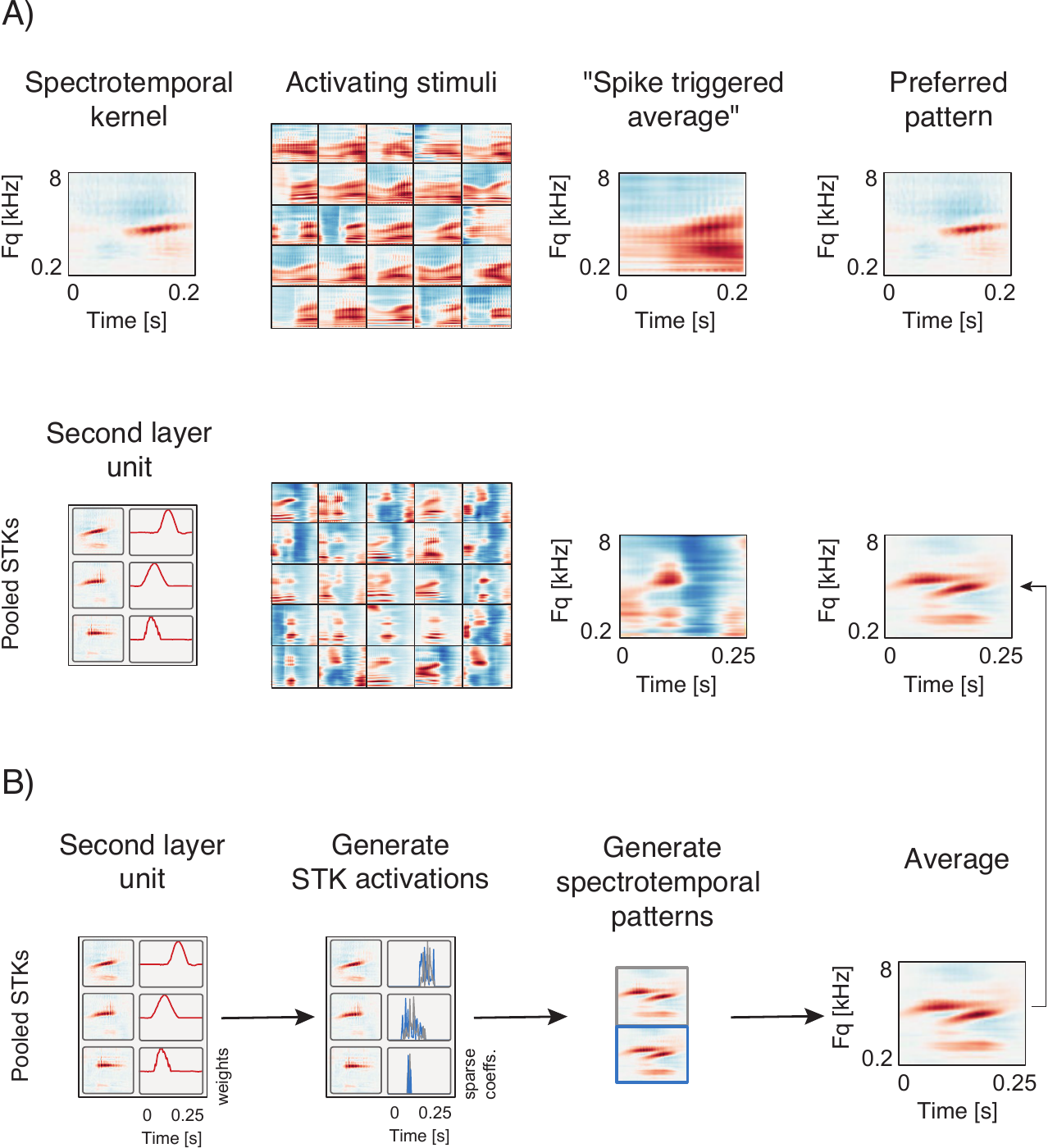}
  \caption{{\bf Estimation of spectrotemporal receptive fields of model units.} A) Top: an example first layer STK (leftmost column) and stimuli which strongly activate it (second column from the left). The "spike triggered average" i.e., the mean of strongly activating stimuli (third column from the left) differs strongly from the true underlying kernel (rightmost column) due to stimulus correlations. Bottom: same as top row for an example second layer unit. Second column from the left depicts strongly activating positive stimuli. Spectrotemporal pattern in rightmost column is derived from the procedure shown in (B). B) Generative estimation of second layer spectrotemporal receptive fields. To avoid the influence of stimulus correlations, we generate a number of STK activation trajectories (second column from the left; gray and blue lines plot two different sets of sampled trajectories) from a distribution encoded by a second-layer unit of interest (leftmost column). For each set of sampled trajectories, we convolve the trajectories with the corresponding STK and sum the results to yield a sampled spectrotemporal pattern ($2$ example patterns are shown in the third column from the left, resulting from the gray and blue samples, respectively). The set of such spectrotemporal patterns are then averaged to estimate the "true" receptive field.}
\label{figAppendixB} 
\end{figure}

Fig. \ref{figAppendixB} A illustrates spectrotemporal receptive field estimation of model units through averaging of strongly activating stimuli (akin to computing spike-triggered average), and the biases that result.

\pagebreak

The "true" spectrotemporal pattern encoded by a STK is simply the STK itself. To identify average spectrotemporal pattern encoded by a second-layer unit we take advantage of the fact that our model is generative (Fig. \ref{figAppendixB}). For each unit we generated a number of STK coefficient trajectories (second panel from the left, Fig. \ref{figAppendixB} B) from the distribution encoded by that particular unit (e.g. leftmost panel, Fig. \ref{figAppendixB} B). We then generated spectrotemporal patterns by convolving the trajectories with the corresponding STKs and summing the results together (see Fig.2). Due to the stochasticity of the generated STK trajectories, some variability is visible among the generated spectrotemporal patterns (third panel from the left, Fig.\ref{figAppendixB} B). We then averaged these patterns to form an estimate of a spectrotemporal receptive field of a second layer unit (right panel, Fig. \ref{figAppendixB} B). Because the estimation process does not involve averaging stimuli, we avoid biasing the estimate of the average preferred spectrotemporal pattern with stimulus correlations.

It is apparent that "spike-triggered averages" (third column, Fig. \ref{figAppendixB} A and B) differ from the true stimulus representation in the model for both layers (as embodied by the STK or the average spectrotemporal pattern generated by a second-layer unit). The discrepancy is due to presence of strong correlations in natural stimuli.

\subsubsection*{Feature Selectivity Index}

For comparison with neural data, we computed the feature selectivity index (FSI) proposed in \cite{MillerFSI}. The FSI is a number lying in the $[0, 1]$ interval.  FSI values close to $1$ imply that stimuli eliciting a response of a neuron (or, in our case, a model unit) are similar to each other (specifically, the stimuli are close to the mean stimulus eliciting a response). When the FSI value is close to $0$, the corresponding neuron spikes at random i.e. stimuli preceding spikes are uncorrelated with the spike-triggered average. The relevance of the FSI for our purposes is that a neuron or model unit that exhibits invariance to some type of stimulus variation should have an FSI less than $1$. By comparing the FSI across layers we hoped to quantify differences in the degree of representational abstraction. 

The FSI computation procedure are described in detail in \cite{MillerFSI}. The only discrepancy between the use of the FSI here and its prior use in neurophysiological studies is that units in our model are continuously active, and do not discretely spike. We emulated the selection of stimuli eliciting spikes by selecting the stimulus excerpts yielding the highest activation of model layer units. In the first layer, we selected the $25$ stimuli yielding highest positive activation. In the second layer, we separately computed FSI indices for stimuli eliciting positive and negative responses, using $25$ stimuli per unit in each case. We then averaged the indices obtained for the two sets of stimuli for each unit.

Computing FSI for an $i$-th unit consists of the following steps:
\begin{enumerate}
\item Compute average of strongly activating stimuli (analogous to spike-triggered average - STA) - separately for positive and negative stimuli in the case of second-layer units.
\item Compute correlations $c_{S,i,n}$ between each strongly activating stimulus $n$ and its respective STA.
\item Compute correlations $c_{R,i,n}$ between the STA and a randomly selected subset of stimuli $n$.
\item Compute the area $A_{S,i}$ under the empirical cumulative distribution function of correlations $c_{S,i}$, $A_{S,i} = \int_{-1}^1 ECDF(c_{S,i})d c_{S,i}$
\pagebreak
\item Compute area $A_{R,i}$ under the empirical cumulative distribution function of $A_{R,i} = \int_{-1}^1 ECDF(c_{R,i}) d c_{R_i}$
\item The FSI of each unit is defined as: $FSI_i = \frac{(A_{R,i} - A_{S,i})}{A_{R,i}}$
\end{enumerate}
\pagebreak

\section*{Appendix C}
\label{AppendixC}
\renewcommand{\thefigure}{C\arabic{figure}}
\setcounter{figure}{0}

\begin{figure}[H]
  \centering
  \includegraphics[scale=.7]{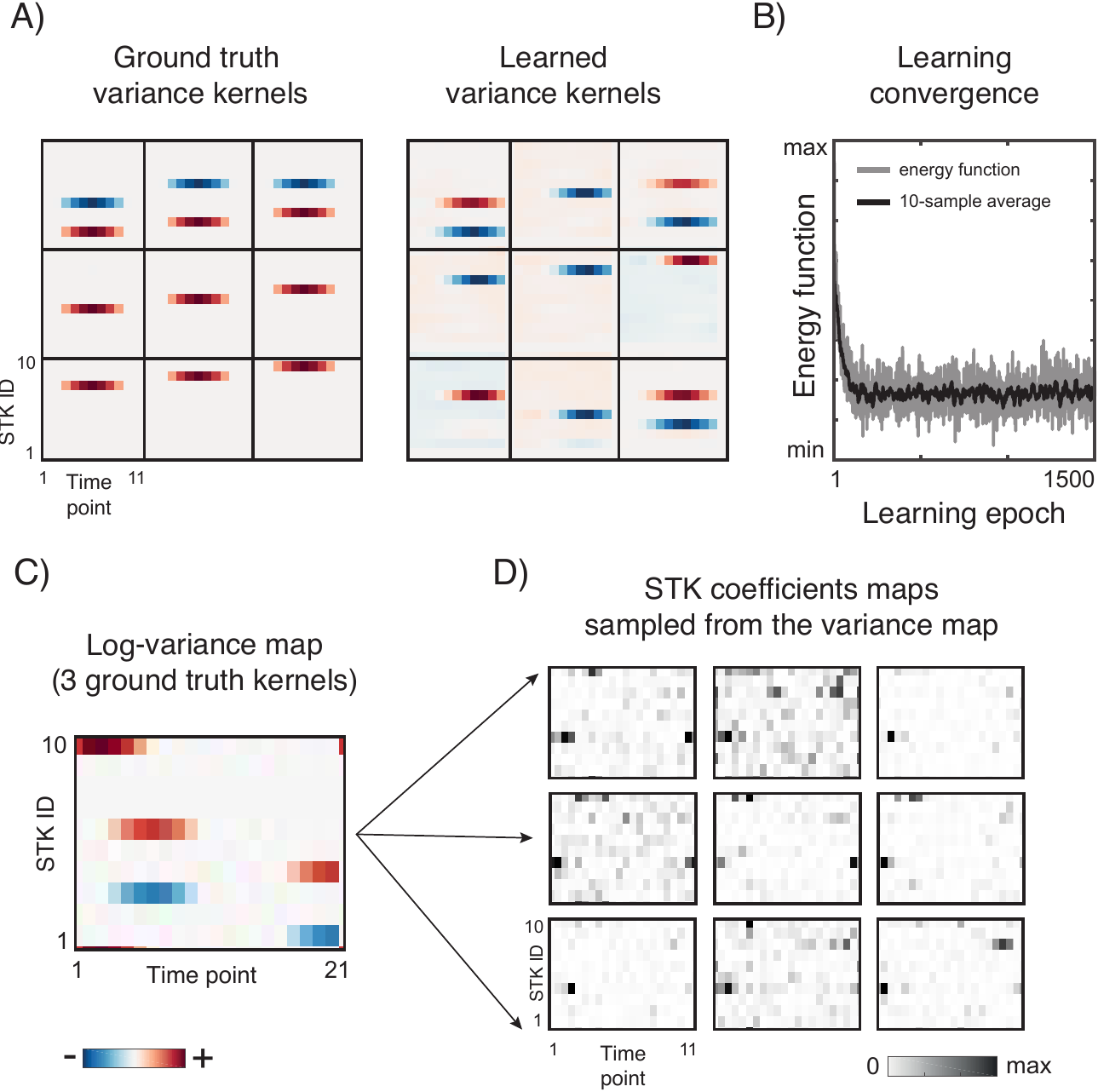}
  \caption{{\bf The model recovers ground truth second-layer kernels from a training dataset. } A) Left panel - nine ground truth second-layer kernels used to generate the toy dataset. The three kernels in the top row are "opponent". Right panel - the generative second-layer kernels recovered by the model. They match the ground truth kernels up to permutation, sign change and time shifts. B) Example variance map generated by the model from the second-layer kernels. Such maps result from superimposing second-layer kernels with coefficients sampled from their sparse prior. This example was generated from three second-layer kernels from panel (A). The variance map encodes temporal changes in the distribution of STK activations. The toy training dataset was generated from 1500 such sampled maps (one per iteration of the learning algorithm). C) STK activations sampled from the variance map in panel B. In the actual training procedure, a single sample was used for each sampled variance map, but here we show $9$ examples to illustrate the stochastic nature of the procedure. D) The convergence of the learning algorithm over $1500$ learning epochs. The gray line depicts values of the second-layer energy function (see Eq.\ref{eq7:E2}). The black line is a $10$-epoch moving average. }
\label{figAppendixC} 
\end{figure}

In order to verify the correctness of the learning algorithm, we generated a toy dataset using 9 second-layer kernels, out of which $3$ were "opponent"  (Fig.\ref{figAppendixC} A, left panel). We generated sample STK activations by randomly superimposing the second-layer kernels to create a "variance map" (Fig.\ref{figAppendixC} B), from which we sampled STK trajectories (Fig.\ref{figAppendixC} C). Each iteration of the learning procedure made a gradient step on the model parameters using a single sample of STK trajectories (from a single sampled variance map). After $1500$ iterations (Fig.\ref{figAppendixC} D), the model converged on the solution visualized in the right panel of Fig.\ref{figAppendixC} A. The generative kernels were recovered up to permutation, sign changes and in some cases, a temporal shift. These latter differences are expected due to the inherent arbitrariness of positive and negative signs in the model and the shift-invariance inherent to a convolutional code. These results demonstrate that the learning algorithm is capable of converging to the correct solution when the data is well-described by the model.

\section*{Acknowledgments}
This material is based upon work supported by the Center for Brains, Minds and Machines (CBMM), funded by NSF STC award CCF-1231216, and by a McDonnell Scholar Award to JHM. The authors thank the members of the McDermott lab for helpful comments on earlier drafts of the manuscript.

%\bibliographystyle{apalike}
%\bibliography{bibliography}

\end{document}